\documentclass{aa}
\usepackage{txfonts}
\usepackage[dvips]{graphicx}
\usepackage{amsfonts}
\usepackage{amssymb}
\usepackage{multirow}
\usepackage{rotating}
\usepackage{natbib}
\bibpunct{(}{)}{;}{a}{}{,}

\newcommand{\te}{$T_{\rm eff}$}
\newcommand{\logg}{$\log{g}$}
\newcommand{\vsini}{$v\sin{i}$}
\newcommand{\kms}{km\,s$^{-1}$}

\newcommand{\gssp}{{\sc gssp}}
\newcommand{\gssps}{{\sc gssp\_single}}
\newcommand{\gsspb}{{\sc gssp\_binary}}
\newcommand{\gsspc}{{\sc gssp\_composite}}

\begin{document}
\title{Grid Search in Stellar Parameters: a software for spectrum analysis of single stars and binary systems\thanks{The \gssp\ software package can be downloaded from https://fys.kuleuven.be/ster/meetings/binary-2015/gssp-software-package}}
\author{A. Tkachenko\inst{1}$^,$\thanks{Postdoctoral Fellow of the Fund for Scientific Research (FWO), Flanders, Belgium}} \institute{Instituut voor Sterrenkunde, KU
Leuven, Celestijnenlaan 200D, B-3001 Leuven, Belgium\\
\email{Andrew.Tkachenko@ster.kuleuven.be}}
\date{Received date; accepted date}
\abstract{The currently operating space missions, as well as those
that will be launched in the near future, (will) deliver
high-quality data for millions of stellar objects. Since the
majority of stellar astrophysical applications still (at least
partly) rely on spectroscopic data, an efficient tool for the
analysis of medium- to high-resolution spectroscopy is needed.}{We
aim at developing an efficient software package for the analysis of
medium- to high-resolution spectroscopy of single stars and those in
binary systems. The major requirements are that the code has a high
performance, represents the state-of-the-art analysis tool, and
provides accurate determinations of atmospheric parameters and
chemical compositions for different types of stars.}{We use the
method of atmosphere models and spectrum synthesis, which is one of
the most commonly used approaches for the analysis of stellar
spectra. Our Grid Search in Stellar Parameters (\gssp) code makes
use of the Message Passing Interface (OpenMPI) implementation, which
makes it possible to run in parallel mode. The method is first
tested on the simulated data and is then applied to the spectra of
real stellar objects.}{The majority of test runs on the simulated
data were successful in the sense that we could recover the
initially assumed sets of atmospheric parameters. We experimentally
find the limits in signal-to-noise ratios of the input spectra,
below which the final set of parameters gets significantly affected
by the noise. Application of the \gssp\ package to the spectra of
three $Kepler$ stars, KIC\,11285625, KIC\,6352430, and KIC\,4931738,
was also largely successful. We found an overall agreement of the
final sets of the fundamental parameters with the original studies.
For KIC\,6352430, we found that dependence of the light dilution
factor on wavelength cannot be ignored, as it has significant impact
on the determination of the atmospheric parameters of this binary
system.}{The \gssp\ software package is a compilation of three
individual program modules suitable for spectrum analysis of single
stars and individual binary components. The code is highly
performant and can be used for spectrum analysis of large samples of
stars.}  \maketitle

\section{Introduction}

Nowadays, spectrum analysis is the main source of precise
atmospheric parameters and chemical compositions of stars. The
problem of chemical composition determination in different classes
of stars is directly linked to the problems of production and
evolution of chemical elements, stellar and galactic evolution, etc.
Interpretation of the observed atmospheric chemical composition in a
star gives a general impression about its evolutionary status. For
example, it is well known that a star initially has the chemical
composition of the molecular cloud it has been formed from. Shortly
after the start of nuclear fusion in the star, the chemical
composition in its central parts undergoes certain changes. At this
stage, the atmospheric composition remains quite stable. On the
main-sequence, certain processes may occur in the star that will
cause the exchange of matter between the stellar interior and the
atmosphere. Examples are the products of the CNO-cycle in the
atmospheres of massive O-B stars
\citep[e.g.,][]{Maeder1983a,Maeder1983b,Massey2003}, or lithium
depletion in the atmospheres of F-K dwarf stars
\citep[e.g.,][]{Soderblom2015,Andrassy2015,Delgado2015}. Some of the
processes that can cause the above mentioned mixing of the material
are convection or turbulent diffusion.

At early stages of giant evolution, the stars generally go through
the phase of deep convective mixing which has certain impact on the
atmospheric composition of light (and later on heavier) chemical
elements. Detailed studies of atmospheric chemical composition can
also reveal signatures of mass loss in massive stars
\citep[e.g.,][]{Chiosi1986} and episodes of mass transfer in close
binary systems \citep[e.g.,][]{Berdyugina1994}. Chemically peculiar
stars are another example of how detailed studies of atmospheric
chemical composition allow for a better understanding of different
physical processes occurring in stars
\citep[e.g.,][]{Pohnl2003,Kochukhov2006}.

Thanks to the launches of several space missions such as MOST
\citep{Walker2003}, CoRoT \citep{Auvergne2009}, and $Kepler$
\citep{Gilliland2010}, the astrophysical fields such as, e.g.,
asteroseismology and exoplanetary science has been revolutionized
during the past decade. High quality photometric data obtained from
space revealed countless number of interesting physical effects in
different types of stellar objects and led to very interesting
discoveries. Needless to say, however, that both fields still
heavily depend on high-quality, high-resolution spectroscopic data,
and a significant fraction of the analyses rely on the
interpretation of combined space-based photometric and ground-based
spectroscopic data. The recently launched Gaia mission
\citep{Perryman2001} and missions such as TESS \citep{Ricker2014}
and PLATO2.0 \citep{Rauer2014} that will be launched in the near
future, will provide high-quality data for a few million of stellar
objects suitable for spectroscopic follow-up from the ground. As
such, efficient and fast tools are particularly needed for the
analysis of ground-based spectroscopic data of single and multiple
stellar objects.

Nowadays, the method of atmosphere models and spectrum synthesis is
a dominant approach in the analysis of high-resolution spectra of
single stars and those in binary systems. One of the advantages of
the spectrum synthesis over the traditional methods that rely on the
calculation of equivalent widths is that the effects of line
blending can be accurately taken into account. This in turn
minimizes the uncertainties in the fundamental stellar parameters
and chemical abundances determination. One example of widely used
software for spectrum analysis of single stars based on the spectrum
synthesis method is the Spectroscopy Made Easy
\citep[SME,][]{Valenti1996} package.

In this paper, we present a new Grid Search in Stellar Parameters
(\gssp) software package for spectrum analysis of high-resolution
spectra of single stars and those in binary system. In the next
sections, we describe the implemented methodology and test the code
both on simulated and real stellar spectra.

\section{Methodology}

In this section we discuss in all necessary detail the methodology
implemented in the \gssp\ software package. This includes the
description of each of the three program modules and the results of
the test runs. In summary, the \gssps\ module is designed for the
analysis of single star spectra and the disentangled light diluted
spectra of the individual components of double-lined spectroscopic
binary system, assuming wavelength-independent light dilution
factor. In this case, the individual binary components are
interpreted independent of each other, just as they were single
stars, unlike the method implemented in the \gsspb\ module which is
specifically designed for simultaneous interpretation of the binary
components' disentangled spectra, by taking into account wavelength
dependence of their light dilution factors. A similar methodology is
implemented in the \gsspc\ module, where the fitting of the
disentangled spectra is replaced by the fitting of the observed
composite spectra, and the wavelength dependence of light dilution
is also assumed.

\begin{table} \tabcolsep 1.0mm\caption{\small
Stellar atmosphere models computed with the {\sc LLmodels} code for
$\xi$\,=\,2\,\kms. The table is re-produced from
\citet{Tkachenko2012}}
\begin{tabular}{lclclc}
\hline\hline
\multicolumn{6}{c}{Parameter, step width\rule{0pt}{9pt}}\\
\hline \multicolumn{1}{c}{$[M/H]$\rule{0pt}{9pt}} & $\Delta[M/H]$
&\multicolumn{1}{c}{\te(K)}
   & $\Delta$\te(K) & \multicolumn{1}{c}{\logg} & $\Delta$\logg\\
\hline --0.8\,--\,+0.8 & 0.1 & \begin{tabular}{l}
                       ~~4\,500\,--\,10\,000\rule{0pt}{9pt}\\ 10\,000\,--\,22\,000\\
                       \end{tabular}
                     & \begin{tabular}{c}
                       100\rule{0pt}{9pt}\\ 250\\
                       \end{tabular}
                     & \begin{tabular}{c}
                       2.5\,--\,5.0\rule{0pt}{9pt}\\ 3.0\,--\,5.0\\
                       \end{tabular}
                     & 0.1\\
\hline
\multicolumn{5}{l}{Total number of models:\rule{0pt}{9pt}} & \textbf{41\,888}\\
\hline
\end{tabular}
\label{Tab: models}
\end{table}

\subsection{Basic methodology}

Although there are small differences in the realization of the
individual algorithms, the general methodology is the same for all
three modules. The software package is based on a grid search in the
fundamental atmospheric parameters and (optionally) individual
chemical abundances of the star (or binary stellar components) in
question. We use the method of atmosphere models and spectrum
synthesis, which assumes a comparison of the observations with each
theoretical spectrum from the grid. For calculation of synthetic
spectra, we use the {\sc SynthV} LTE-based radiative transfer code
\citep{Tsymbal1996} and a grid of atmosphere models precomputed with
the {\sc LLmodels} code \citep{Shulyak2004}. Our grid of models
covers large ranges in all fundamental atmospheric parameters and is
freely distributed together with the software package itself. The
summary of the available models is given in Table~\ref{Tab: models}.
In fact, the \gssp\ package is compatible with any kind of
atmosphere model grid as long as the models are provided in the
Kurucz format. The authors also possess a grid of Kurucz
models\footnote{http://kurucz.harvard.edu/grids.html}
\citep{Kurucz1993} which have been interpolated in all fundamental
parameters to match the resolution of our {\sc LLmodels} grid. The
interpolated grid of Kurucz models is available upon request.

We allow for optimization of five stellar parameters at a time:
effective temperature \te, surface gravity \logg, metallicity [M/H],
microturbulent velocity $\xi$, and projected rotational velocity
\vsini\ of the star. The synthetic spectra can be computed in any
number of wavelength ranges, and each considered spectral interval
can be from a few angstroem up to a few thousand angstroem wide. As
long as the global metallicity of the star is determined/known, the
[M/H] parameter can be replaced in the grid by the abundance of an
arbitrary chemical element. The individual abundances have to be
iterated element by element, thus there is no option to optimize
abundances of more than one element at the same time. From our
experience, any reasonable deviations of the individual abundances
from the global atmospheric metallicity (within about 0.5~dex) have
little (within typical 1$\sigma$ error bars) to no influence on
abundances of other chemical elements. An exception needs to be made
however for the elements showing a large amount of lines in their
spectrum: should such an element be found to show significant
over/underabundance, all chemical element abundances iterated before
must be re-determined, taking into account the fact that one of the
dominant chemical elements in the spectrum shows peculiarities. For
this reason, we advice the user to start with chemical elements
having the largest amount of lines in the spectrum when determining
the detailed chemical composition of the star in question.

The grid of theoretical spectra is built from all possible
combinations of the above mentioned parameters. Each spectrum from
the grid is compared to a priori normalized observed spectrum of the
star and the $\chi^2$ merit function is used to judge the goodness
of fit. The code delivers the set of best fit parameters, the
corresponding synthetic spectrum, and the ASCII file containing the
individual parameter values for all grid points and the
corresponding $\chi^2$ values. The $\chi^2$ we deliver is the
reduced $\chi^2$, that is the $\chi^2$ value normalized to the
number of pixels across the observed spectrum minus the number of
free parameters. Based on the $\chi^2$ statistics and assuming
normal distribution, we also compute 1$\sigma$ uncertainty level in
terms of $\chi^2$ and output this value along with the $\chi^2$
distributions. We also account for possible global-scale
imperfection in the normalization of the observed spectrum by means
of a scaling factor that is computed from the least-squares fit of
the synthetic spectrum to the observations and is applied to the
latter. The value of the scaling factor is provided in the final
$\chi^2$ table along with the other grid search parameters.

\begin{figure}[t]
\centering
\includegraphics[scale=0.45]{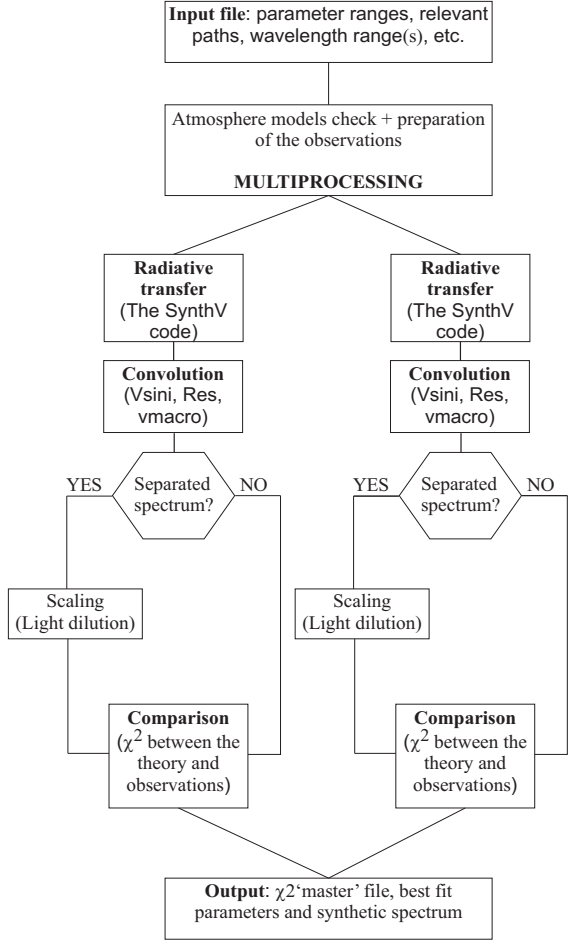}
\caption{{\small Diagram illustrating the algorithm implemented in
the \gssps\ software module. The two identical branches showed in
the diagram indicate multiprocessing. See text for more details.}}
\label{Fig: gssps_diagram}
\end{figure}

For the analysis of binary stars, whether the disentangled spectra
or the observed composite spectra are used for the characterization
of the system, the percentage contribution of a stellar component to
the total light of the system becomes one of the key parameters.
This light contribution is often referred to as a {\it light
dilution factor} (designated as $f_i$ throughout this paper) and
effects the depths of lines in the spectra of both binary
components. This correction factor for the line depths has to be
taken into account in the analysis, and is ideally determined along
with the other atmospheric parameters of the star. There are two
basic ways of accounting for the light dilution factor in the
spectroscopic analysis: i) assuming a constant scaling of the line
depths across the entire wavelength range (unconstrained fitting of
the disentangled spectra), and ii) taking into account
wavelength-dependence of the light dilution due to possible
difference in spectral type and/or luminosity class for the two
stars forming a binary system \citep[constrained fitting of the
disentangled spectra,][]{Tamajo2011}. Which of the two methods to
use depends on the analysis approach, on whether the individual
binary components are characterized independent of each other or
simultaneously by means of fitting their disentangled spectra. More
details on both methods are given in Sect.~2.2 and 2.3.

The grid search method is generally more CPU time demanding than any
of the optimization algorithms but has a big advantage that it
guarantees that the global minimum solution will be found as long as
the considered parameter range is large enough. We use the {\sc
OpenMPI\footnote{http://www.open-mpi.org/}} distribution to
parallelize our code, which solves the CPU time-related problem and
makes the code fast and highly performant. At least for one of the
software modules, there seems to be a linear relation between the
number of used CPUs and the calculation time. However, there still
exists an upper limit for the number of CPUs to be used. The limit
strongly depends on the configuration of the system and is likely to
occur when the system load becomes large enough to start effecting
the performance of the code. We will provide rough estimates of the
calculation times below, when discussing each of the \gssp\ package
modules individually.

\subsection{\gssps\ module}

\begin{figure}[t]
\includegraphics[scale=0.92]{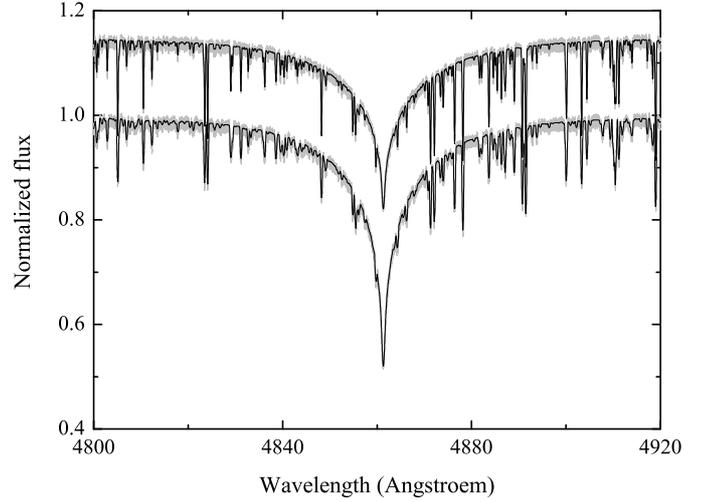}
\caption{{\small Quality of the fit of the simulated data of
KIC\,11285625 (model 1b in Table~\ref{Tab: unconstrained_fitting})
with the \gssps\ software package. The light dilution factor is
0.6/0.4 for the primary/secondary component. The simulated observed
spectrum is shown with light grey lines, the black solid line refers
to the best fit synthetic spectrum. The spectrum of the secondary
was vertically shifted by a constant value for better
visualization.}} \label{Fig: KIC11285625_simulated}
\end{figure}

In this particular module we treat the observed spectrum as the one
of a single star. The module is thus applicable to spectra of single
stellar objects as well as to the disentangled spectra of multiple
stellar systems. In the latter case, we assume that the light of the
star in question is diluted by a certain wavelength-independent
factor, and the disentangled spectrum of a star is represented as
follows
\begin{equation}
R_i ^{\rm dis}=\frac{F_{\nu,i}+\sum\limits_{j=1}^{N-1}
F_{\nu,j}^c}{\sum\limits_{k=1}^N F_{\nu,k}^c}, \ \ \ \ \ \
F_{\nu}=\pi I_{\nu}\left(\frac{\Re}{d}\right)^2 \label{Eg:
disentangled_spectrum}
\end{equation}
where $I_{\nu}$ is the specific intensity of the star at a given
wavelength/frequency, $\Re$ is the radius of the star, and $d$ is
the distance to the object. After a few trivial mathematical
operations, for either of the components of the binary system we
obtain in terms of line depths
\begin{equation}
r_i^{\rm dis}=r_i^{\rm th}\frac{1}{1+\alpha}=r_i^{\rm th}f_i, \ \ \
\ \ \ \alpha=\frac{I_{\nu,i}^c}{I_{\nu,{\rm
comp}}^c}\left(\frac{\Re_i}{\Re_{\rm comp}}\right)^2 \label{Eg:
diluted_spectrum}
\end{equation}
where superscripts ``th'' and ``c'' refer to the theoretical
spectrum and continuum intensity, respectively, and subscript
``comp'' points to the companion star. The factor $f_i=1/(1+\alpha)$
is the light dilution factor briefly outlined in Sect.~2.1. In this
particular case of interpreting the individual binary components'
spectra independent of each other, nothing is known about the
continuum intensity ratio of the two stars and the light dilution
factor $f_i$ is assumed to be wavelength independent. In practice,
each synthetic spectrum from the computed grid is represented in the
form of Eq.~\ref{Eg: diluted_spectrum} and compared to the observed
disentangled spectrum on the scale of the latter.

\begin{table}
\tabcolsep 1.2mm\caption{Results of the spectrum analysis of both
stellar components of KIC\,11285625 with the \gssps\ and \gsspb\
software packages (see text for details). The parameters derived by
\citet{Debosscher2013} are given as a reference; the \logg\ values
and the radii radio were derived by the authors from light curve
solution.}\label{Tab: KIC11285625}
\begin{tabular}{llllll} \hline\hline
 \multirow{2}{*}{Parameter} & \multirow{2}{*}{Unit} &
 \multicolumn{2}{c}{\gssps} & \multicolumn{2}{c}{\gsspb}\\
  & & \multicolumn{1}{c}{primary} & \multicolumn{1}{c}{secondary}& \multicolumn{1}{c}{primary} & \multicolumn{1}{c}{secondary}\\\hline
\te & K & 7\,200$^{+80}_{-78}$ & 7\,300$^{+100}_{-100}$ & 6\,980$^{+82}_{-80}$\rule{0pt}{9pt} & 7\,100$^{+175}_{-130}$\vspace{1mm}\\
\logg & dex & 3.91$^{+0.21}_{-0.21}$ & 4.18 (fixed) & 3.82$^{+0.25}_{-0.25}$ & 4.18 (fixed)\vspace{1mm}\\
$\xi$ & \kms & 1.19$^{+0.16}_{-0.14}$ & 0.0$^{+0.6}$ & 0.72$^{+0.51}_{-0.41}$ & 0.10$^{+0.40}$\vspace{1mm}\\
\vsini & \kms & 14.5$^{+1.0}_{-0.9}$ & 8.6$^{+1.0}_{-0.9}$ & 15.6$^{+1.2}_{-1.2}$ & 8.9$^{+1.3}_{-1.3}$\vspace{1mm}\\
${\rm [M/H]}$ & dex & --0.25$^{+0.08}_{-0.08}$ & --0.21$^{+0.10}_{-0.10}$ & --0.43$^{+0.09}_{-0.09}$ & --0.37$^{+0.15}_{-0.15}$\vspace{1mm}\\
R$_1$/R$_2$ ($f_i$) & & 0.64$^{+0.05}_{-0.05}$ &
0.27$^{+0.02}_{-0.02}$ &
\multicolumn{2}{c}{1.62$^{+0.16}_{-0.15}$}\vspace{1mm}\\ \hline
\multicolumn{6}{c}{{\bf Parameters from \citet{Debosscher2013}\rule{0pt}{11pt}}}\\
 & & \multicolumn{2}{c}{primary\rule{0pt}{9pt}} & \multicolumn{2}{c}{secondary}\vspace{1mm}\\
\te & K & \multicolumn{2}{c}{6\,960$^{+100}_{-100}$} & \multicolumn{2}{c}{7\,195$^{+200}_{-200}$}\vspace{1mm}\\
\logg & dex & \multicolumn{2}{c}{3.97 (fixed)} & \multicolumn{2}{c}{4.18 (fixed)}\vspace{1mm}\\
$\xi$ & \kms & \multicolumn{2}{c}{0.95$^{+0.30}_{-0.30}$} & \multicolumn{2}{c}{0.09$^{+0.25}_{-0.25}$}\vspace{1mm}\\
\vsini & \kms & \multicolumn{2}{c}{14.2$^{+1.5}_{-1.5}$} & \multicolumn{2}{c}{8.4$^{+1.5}_{-1.5}$}\vspace{1mm}\\
${\rm [M/H]}$ & dex & \multicolumn{2}{c}{--0.49$^{+0.15}_{-0.15}$} & \multicolumn{2}{c}{--0.37$^{+0.30}_{-0.30}$}\vspace{1mm}\\
R$_1$/R$_2$ & & \multicolumn{4}{c}{1.44$^{+0.02}_{-0.02}$}\vspace{1mm}\\
\hline \multicolumn{6}{l}{Note: dilution factors $f_i$ are reported
for the \gssps\ mode \rule{0pt}{9pt}}
\end{tabular}
\end{table}

\begin{table}
\tabcolsep 1.2mm\caption{Same as Table~\ref{Tab: KIC11285625} but
for KIC\,6352430. The reference parameters are those from
\citet{Papics2013}.}\label{Tab: KIC6352430}
\begin{tabular}{llllll} \hline\hline
 \multirow{2}{*}{Parameter} & \multirow{2}{*}{Unit} &
 \multicolumn{2}{c}{\gssps} & \multicolumn{2}{c}{\gsspb}\\
  & & \multicolumn{1}{c}{primary} & \multicolumn{1}{c}{secondary}& \multicolumn{1}{c}{primary} & \multicolumn{1}{c}{secondary}\\\hline
\te & K & 13\,200$^{+100}_{-100}$ & 7\,065$^{+100}_{-100}$ & 13\,060$^{+115}_{-115}$\rule{0pt}{9pt} & 7\,340$^{+240}_{-240}$\vspace{1mm}\\
\logg & dex & 4.07$^{+0.04}_{-0.04}$ & 3.70$^{+0.20}_{-0.20}$ & 4.03$^{+0.05}_{-0.05}$ & 3.60$^{+0.56}_{-0.56}$\vspace{1mm}\\
$\xi$ & \kms & 1.12$^{+0.53}_{-0.58}$ & 3.75$^{+0.50}_{-0.50}$ & 1.90$^{+0.43}_{-0.41}$ & 4.50$^{+0.90}_{-0.90}$\vspace{1mm}\\
\vsini & \kms & 72.8$^{+4.1}_{-3.8}$ & 8.5$^{+1.0}_{-0.9}$ & 71.4$^{+3.2}_{-3.2}$ & 7.0$^{+2.3}_{-2.2}$\vspace{1mm}\\
${\rm [M/H]}$ & dex & --0.08$^{+0.05}_{-0.05}$ & --0.24$^{+0.10}_{-0.10}$ &--0.19$^{+0.06}_{-0.06}$ & 0.01$^{+0.18}_{-0.18}$\vspace{1mm}\\
R$_1$/R$_2$ ($f_i$) & & 0.95$^{+0.01}_{-0.01}$ & 0.06$^{+0.01}_{-0.01}$ & \multicolumn{2}{c}{1.86$^{+0.11}_{-0.11}$}\vspace{1mm}\\
\hline
\multicolumn{6}{c}{{\bf Parameters from \citet{Papics2013}\rule{0pt}{11pt}}}\\
 & & \multicolumn{2}{c}{primary\rule{0pt}{9pt}} & \multicolumn{2}{c}{secondary}\vspace{1mm}\\
\te & K & \multicolumn{2}{c}{12\,810$^{+200}_{-200}$} & \multicolumn{2}{c}{6\,805$^{+100}_{-100}$}\vspace{1mm}\\
\logg & dex & \multicolumn{2}{c}{4.05$^{+0.05}_{-0.05}$} & \multicolumn{2}{c}{4.26$^{+0.15}_{-0.15}$}\vspace{1mm}\\
$\xi$ & \kms & \multicolumn{2}{c}{2.0 (fixed)} & \multicolumn{2}{c}{2.0 (fixed)}\vspace{1mm}\\
\vsini & \kms & \multicolumn{2}{c}{69.8$^{+2.0}_{-2.0}$} & \multicolumn{2}{c}{9.8$^{+1.0}_{-1.0}$}\vspace{1mm}\\
${\rm [M/H]}$ & dex & \multicolumn{2}{c}{--0.13$^{+0.07}_{-0.07}$} & \multicolumn{2}{c}{--0.33$^{+0.10}_{-0.10}$}\vspace{1mm}\\
\hline \multicolumn{6}{l}{Note: dilution factors $f_i$ are reported
for the \gssps\ mode \rule{0pt}{9pt}}
\end{tabular}
\end{table}

\begin{table}
\tabcolsep 1.2mm\caption{Same as Table~\ref{Tab: KIC11285625} but
for KIC\,4931738. The reference parameters are those from
\citet{Papics2013}.}\label{Tab: KIC4931738}
\begin{tabular}{llllll} \hline\hline
 \multirow{2}{*}{Parameter} & \multirow{2}{*}{Unit} &
 \multicolumn{2}{c}{\gssps} & \multicolumn{2}{c}{\gsspb}\\
  & & \multicolumn{1}{c}{primary} & \multicolumn{1}{c}{secondary}& \multicolumn{1}{c}{primary} & \multicolumn{1}{c}{secondary}\\\hline
\te & K & 14\,120$^{+235}_{-225}$ & 11\,975$^{+245}_{-240}$ & 14\,150$^{+320}_{-320}$\rule{0pt}{9pt} & 11\,780$^{+285}_{-280}$\vspace{1mm}\\
\logg & dex & 4.04$^{+0.10}_{-0.09}$ & 4.40$^{+0.11}_{-0.11}$ & 4.15$^{+0.07}_{-0.09}$ & 4.50$^{+0.16}_{-0.16}$\vspace{1mm}\\
$\xi$ & \kms & 0.40$^{+0.90}_{-0.40}$ & 0.88$^{+0.75}_{-0.88}$ & 0.00$^{+1.10}$ & 0.55$^{+1.45}$\vspace{1mm}\\
\vsini & \kms & 11.3$^{+1.2}_{-1.2}$ & 6.5$^{+1.2}_{-1.2}$ & 11.2$^{+2.2}_{-2.1}$ & 6.6$^{+1.8}_{-1.8}$\vspace{1mm}\\
${\rm [M/H]}$ & dex & --0.15$^{+0.09}_{-0.10}$ & 0.06$^{+0.09}_{-0.10}$ &--0.19$^{+0.21}_{-0.21}$ & 0.14$^{+0.15}_{-0.15}$\vspace{1mm}\\
R$_1$/R$_2$ ($f_i$) & & 0.72$^{+0.03}_{-0.03}$ &
0.34$^{+0.02}_{-0.02}$ & \multicolumn{2}{c}{1.26$^{+0.04}_{-0.03}$}
\vspace{1mm}\\\hline
\multicolumn{6}{c}{{\bf Parameters from \citet{Papics2013}\rule{0pt}{11pt}}}\\
 & & \multicolumn{2}{c}{primary\rule{0pt}{9pt}} & \multicolumn{2}{c}{secondary}\vspace{1mm}\\
\te & K & \multicolumn{2}{c}{13\,730$^{+200}_{-200}$} & \multicolumn{2}{c}{11\,370$^{+250}_{-250}$}\vspace{1mm}\\
\logg & dex & \multicolumn{2}{c}{3.97$^{+0.05}_{-0.05}$} & \multicolumn{2}{c}{4.37$^{+0.10}_{-0.10}$}\vspace{1mm}\\
$\xi$ & \kms & \multicolumn{2}{c}{2.0 (fixed)} & \multicolumn{2}{c}{2.0 (fixed)}\vspace{1mm}\\
\vsini & \kms & \multicolumn{2}{c}{10.5$^{+1.0}_{-1.0}$} & \multicolumn{2}{c}{6.6$^{+1.0}_{-1.0}$}\vspace{1mm}\\
${\rm [M/H]}$ & dex & \multicolumn{2}{c}{--0.24$^{+0.10}_{-0.10}$} &
\multicolumn{2}{c}{+0.10$^{+0.10}_{-0.10}$}\vspace{1mm}\\ \hline
\multicolumn{6}{l}{Note: dilution factors $f_i$ are reported for the
\gssps\ mode \rule{0pt}{9pt}}
\end{tabular}
\end{table}

Figure~\ref{Fig: gssps_diagram} illustrates the multiprocess
algorithm implemented in the \gssps\ software module. In the first
step, the code reads in all necessary information provided by the
user in the configuration file. We provide an example of the input
file with detailed description of each entry in the Appendix. In the
next step, the code sets up the grid and checks whether the required
grid of atmosphere models is available. Should any of the models not
exist, the user will be notified about the need to revise the input
set up, and the detailed report is saved in one of the log-files. At
this point, the observed spectrum can be optionally cross-correlated
with the first synthetic spectrum from the grid, and the radial
velocity (RV) is computed as the first order moment of the
cross-correlation function. This way, the code accounts for a
possible RV shift of the observed spectrum with respect to the
laboratory wavelength, should the user request that. As soon as the
grid has been set up (depending on the grid size, this might take up
to half a minute of time), the code enters a multiprocessing mode
and performs the actual calculations. The {\sc SynthV} code is
executed on each of the available CPUs and provides a synthetic
spectrum for a given set of atmospheric parameters and in the
requested wavelength range. The code computes specific line and
continuum intensities for different positions of the stellar disk,
thus providing an accurate treatment of the limb darkening effect.
As soon as the spectrum is computed, a convolution program is
executed to perform the disk integration and a convolution of the
spectrum with projected rotational and macroturbulent velocities,
and the resolving power of the instrument. The convolution code
outputs the normalized synthetic spectrum as well as the line and
continuum fluxes. The above described calculations may take from a
few seconds up to a few minutes, depending on the spectral type and
considered wavelength range. Once the convolved synthetic spectrum
is released, there are two options for the code to contiue with (see
Fig.~\ref{Fig: gssps_diagram}): immediate comparison with the
observations (the case of the spectrum of a single star), or
dilution according to Eq.~\ref{Eg: diluted_spectrum} and then the
comparison with the observations (the case of the disentangled
spectrum of a binary component). In either case, the final output of
an individual process is the goodness of fit of the observed
spectrum in terms of the $\chi^2$ merit function value, whereas the
computed synthetic spectrum is immediately deleted. As soon as the
calculations are finished on all the CPUs, the main process merges
individual $\chi^2$ tables, finds the minimum in $\chi^2$, computes
the best fit synthetic spectrum, and outputs the set of best fit
parameters. We also provide the $\chi^2$ value corresponding to the
1$\sigma$ uncertainty level computed from $\chi^2$ statistics and
assuming normal distribution. This way, the projection of all data
points on the parameter in question can be computed afterwards from
the $\chi^2$ ``master file'', and the 1$\sigma$ uncertainties are
obtained in a straightforward manner by fitting the $\chi^2$
distribution with a polynomial function and searching for its
intersection points with the 1$\sigma$ level in $\chi^2$ provided by
the \gssps\ module.

The \gssps\ module is reasonably fast and the final solution is
obtained within 5--6 minutes by running the code on 8 CPUs, and
assuming a typical grid size of 2\,000--3\,000 synthetic spectra and
a 1\,000~\AA\ wide wavelength range. The calculations may take a bit
longer when fitting disentangled spectra, as there is an additional
parameter (light dilution factor) to account for. The extra
calculation time typically does not exceed 10\% of the computation
time in a single star spectrum mode, unless the grid resolution in
the dilution factor is made unreasonably high.

\begin{figure}[t]
\includegraphics[scale=0.92]{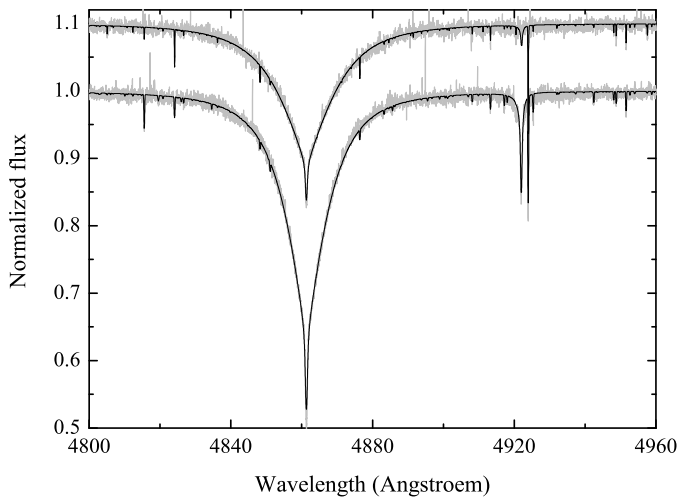}
\caption{{\small Comparison between the observed disentangled
spectra (light grey lines) of both stellar components of the
KIC\,4931738 system and the best fit synthetic spectra (black lines)
computed with the \gssps\ software module. The spectrum of the
secondary was vertically shifted by a constant value for better
visualization.}} \label{Fig: KIC4931738_real}
\end{figure}

The \gssps\ software module is in fact an extension of the original
version of the \gssp\ code \citep{Lehmann2011,Tkachenko2012} to make
it compatible with the disentangled diluted spectra of stellar
components in binary systems. The module was tested on simulated
data of single stars; in addition, we refer the reader to the papers
of \citet{Lehmann2011},
\citet{Tkachenko2012,Tkachenko2013a,Tkachenko2013b},
\citet{Lampens2013}, \citet{Papics2014}, \citet{VanReeth2015} for
several examples of the code application to the spectra of single
stars, including some well studied objects like Vega.

For tests on the light diluted spectra of individual binary
components, we decided not to simulate the spectra of arbitrary
objects but instead used the binary components of KIC\,6352430
\citep{Papics2013} and KIC\,11285625 \citep{Debosscher2013} as
prototype stars. In our simulations, we assumed different dilution
factors as well as different values of S/N ratio; the summary of the
obtained results is given in Table~\ref{Tab: unconstrained_fitting}.
Overall, we could recover the parameters of the input spectra
assumed in all our simulations; the largest deviations, as well as
1$\sigma$ uncertainties, have been observed for low S/N ratio
spectra. The quality of the fit to one pair of spectra is
illustrated in Fig.~\ref{Fig: KIC11285625_simulated}. The only set
up where we have encountered certain difficulties in recovering the
assumed fundamental parameters refers to the case of peculiar stars
(see Table~\ref{Tab: peculiarities_simulated}, the column indicated
as ``\gssps''). In the simulations of KIC\,11285625, we assumed an
over/under-abundance of iron/silicon by 0.3/0.5~dex with respect to
the global metallicity of the star for the primary/secondary
component. For the KIC\,6352430 system, the primary star was assumed
to be peculiar in helium (+0.2~dex compared to the solar
composition), whereas the secondary component was set up as a normal
star with its own global metallicity. Since in the unconstrained
fitting the spectra of individual components are considered
separately, there is no way the results obtained for one of the
components will influence the parameters of the companion. For this
reason, we give no parameters for the secondary of KIC\,6352430 in
Table~\ref{Tab: peculiarities_simulated} as they will be the same as
the ones depicted in the second column of Table~\ref{Tab:
KIC4931738}. We find that when a chemical element dominating the
observed spectrum of the star shows significant over-/underabundance
(iron in the primary of KIC\,11285625), the global metallicity value
gets affected and impacts several other fundamental parameters as
well (e.g., microturbulent velocity and surface gravity). Although
this is not an unexpected result and the majority of the parameter
values are still within 1$\sigma$ uncertainties from the assumed
values, the fact calls for a certain attention.

After testing the code on simulated data, we have applied the
\gssps\ module to the real disentangled spectra of both stellar
components of the KIC\,11285625, KIC\,6352430, and KIC\,4931738
systems. The latter object was also analysed by \citet{Papics2013}
and consists of two stars of similar spectral types, luminosity
classes, and rotation rates. The results of our analyses for all
three binary systems are depicted in Tables~\ref{Tab: KIC11285625},
\ref{Tab: KIC6352430}, and \ref{Tab: KIC4931738}, respectively, in
the columns indicated as ``\gssps''. The parameters obtained by us
are slightly different from those presented in the original studies,
though the majority of the parameters still agree within the
reported 1$\sigma$ uncertainties. The obtained discrepancies for
KIC\,6352430 and KIC\,4931738 may be associated with the fact that
the microturbulent velocity was fixed to 2~\kms\ in the original
study by \citet{Papics2013}, whereas it was set as a free parameter
in our analysis. In addition, \citet{Papics2013} assumed some
dilution factor values coming from the preliminary fit of several
observed composite spectral lines and used those to re-normalize the
disentangled spectra of both binary components to the individual
continua. In our case, both factors were set as free parameters and
optimized along with other fundamental parameters for each of the
components. As an example, in Figure~\ref{Fig: KIC4931738_real} we
show the quality of the fit of the observed disentangled spectra of
both components of the KIC\,4931738 system by the best fit synthetic
spectra computed from the parameters depicted in the first two
columns of Table~\ref{Tab: KIC4931738}.

As it has been mentioned by \citet{Debosscher2013} in the original
study, the spectroscopic data of KIC\,11285625 have low S/N, which
naturally propagates to the quality of the disentangled spectra. The
secondary component, which contribution to the total light amounts
to $\sim$30\%, suffered the most and its disentangled spectrum has
rather low S/N and shows pronounced undulations in the local
continuum. We anyway attempted to fit the spectrum of the secondary
setting \te, \logg, $\xi$, \vsini, and [M/H] as free parameters in
our analysis. We ended up with a clearly unreliable solution, in
particular the value of \logg\ was found to exceed 5.0~dex, which is
not expected for the main-sequence star of spectral type F. Thus,
following the procedure adopted in the original study of
\citet{Debosscher2013}, we fixed surface gravity of the secondary to
the value obtained from the light curve solution, and optimized the
four remaining fundamental parameters. Since the disentangled
spectrum of the primary has higher S/N and suffered less from
artifacts of spectral disentangling, we optimized all five
fundamental parameters for this star, with the particular goal to
compare the spectroscopic \logg\ value with the one obtained from
the light curve solution. The results of the analysis are summarized
in Table~\ref{Tab: KIC11285625} and show good agreement between the
spectroscopic and photometric values of \logg\ for the primary
component, and overall agreement for all parameters of both binary
components with the original study.

\begin{figure}[t]
\centering
\includegraphics[scale=0.45]{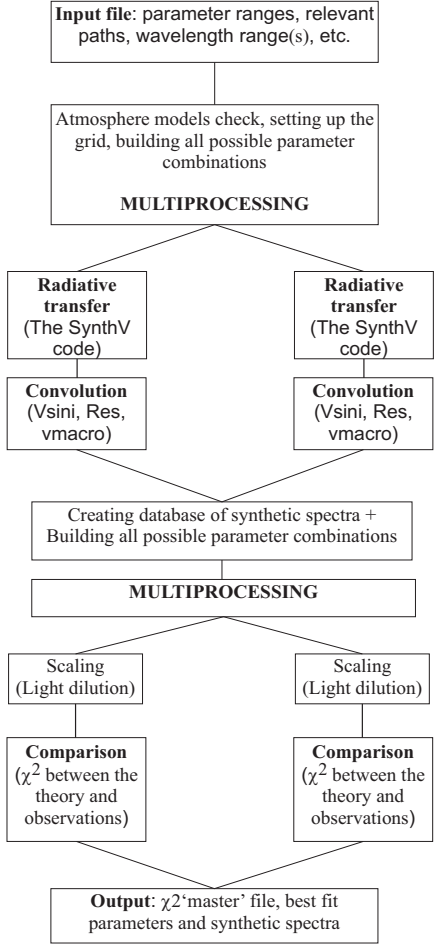}
\caption{{\small Diagram illustrating the algorithm implemented in
the \gsspb\ software module. The two identical branches showed in
the diagram indicate multiprocessing. See text for more details.}}
\label{Fig: gsspb_diagram}
\end{figure}

\subsection{\gsspb\ module}

This module has been specifically designed for fitting disentangled
spectra of both binary components simultaneously. The procedure is
in a sense similar to the one of ``constrained fitting'' suggested
by \citet{Tamajo2011}, were one assumes that the sum of the two
light factors is identical to unity, and thus the change in light
dilution of one of the components affects the amount of diluted
light for the companion star. Instead of assuming
wavelength-independent light dilution for each of the components, we
optimize the ratio of the radii of the binary components which is
obviously the same for both stars. Starting from the definition of
disentangled spectra of individual binary components given in
Eq.~\ref{Eg: disentangled_spectrum}, and dividing all terms on the
right hand side by the product between the continuum intensity of
the secondary and its squared radius, we obtain for the primary and
secondary components of a binary
\begin{equation}
R_1^{\rm dis} = \frac{\alpha R_1^{\rm th} + 1}{1+\alpha};\ \ \ \ \
R_2^{\rm dis} = \frac{\alpha + R_2^{\rm th}}{1+\alpha}, \ \ \ \ {\rm
with}\ \
\alpha=\frac{I_{\nu,1}^c}{I_{\nu,2}^c}\left(\frac{\Re_1}{\Re_2}\right)^2
\label{diluted_spectrum_gsspb}
\end{equation}
or, in terms of line depths
\begin{equation}
r_1^{\rm dis} = r_1^{\rm th}\frac{\alpha}{1+\alpha};\ \ \ \ \
r_2^{\rm dis} = r_2^{\rm th}\frac{1}{1+\alpha}
\end{equation}
Just as in the formulation of \citet{Tamajo2011}, the sum of the two
dilution factors is identical to unity. By optimizing the ratio of
the radii, we take into account the wavelength-dependence of the
light dilution factor through the ratio of continuum intensities of
the two stars. Obviously, any change in the atmospheric parameters
of one of the binary components will influence its continuum
intensity, which will in turn impact the light dilution factor for
the companion star. In practice, disentangled spectra of two binary
components are analyzed simultaneously by scaling synthetic spectra
from the corresponding grids and comparing them to the observations
on the scale of the latter.

Figure~\ref{Fig: gsspb_diagram} visualizes the algorithm implemented
in the \gsspb\ software module. Although the general principle is
the same as for the \gssps\ module, there are a number of
differences that we discuss here in more detail. After reading all
necessary information from the input file and checking the
atmosphere models for availability, the code sets up the grid of
synthetic spectra that need to be computed. As we have to deal with
two components simultaneously, the algorithm of setting up the grid
is different from the one implemented in the \gssps\ module. To
supply the code with maximum efficiency, we first build individual
grids for both components of a binary system, and in the second
step, a unique grid common for both stars is build by excluding any
possible overlap in the parameters between the individual grids.
This step is needed to avoid unnecessary repetitive calculations of
theoretical spectra. For example, in the case of a binary system
with two identical stars and thus identical initial parameter grids,
only the primary's grid will be computed and used for both binary
components afterwards. The individual grids are used at this step to
build all possible combinations of primary and secondary spectra,
which will be processed at a later stage when looking for the best
fit solution. For example, three grid points for each of the binary
components will result in nine combinations to be processed in
total, regardless of any possible overlap between the individual
grids. As soon as the grids were set up (takes less than a minute of
time), the code enters the parallel mode and starts the actual
calculations of synthetic spectra as described in Sect.~2.1.

The normalized synthetic spectra as well as the corresponding
continuum fluxes are stored in a database, which is an unformatted
direct access Fortran~90 file. The advantage of using such a file
over storing the data in an ASCII file is that any I/O operation
with such a file is much faster as there is no conversion from
machine code to a readable format. By storing the data in such a
file we also avoid any possible problems with insufficient RAM
memory, which might occur when keeping the (sometimes very large)
grid of synthetic spectra in the memory. As soon as the database of
spectra is produced, it is first made available to all CPUs involved
in the calculations, and then the code enters the parallel mode for
the second time. At this stage, every pre-build combination of
primary and secondary spectra is processed, which involves dilution
of both synthetic spectra extracted from the database and comparison
with the observations on the scale of the latter. After processing
all possible combinations, the $\chi^2$ files from individual CPUs
are merged into a ``master file'', and the best fit parameters and
synthetic spectra for both binary components are released. Similar
to the \gssps\ software module, the 1$\sigma$ level in $\chi^2$ is
provided so that the corresponding uncertainties in all the
parameters can be computed afterwards.

Despite a very large number of combinations that have to be
processed, the \gsspb\ module is reasonably fast. Assuming the same
set up as the one described in Sect~2.1 - 8 CPUs and a 1\,000~\AA\
wide wavelength range, - and $\sim$500 synthetic spectra in a grid,
the final solution is obtained within 15--20 minutes. To provide an
estimate on the number of combinations, the above mentioned grid of
$\sim$500 spectra coupled with only three grid points in the radii
ratio results in about 200\,000 combinations to be processed by the
code.

The \gsspb\ software module has been extensively tested by us both
on simulated and real spectra of double-lined spectroscopic binary
stars. Similar to the tests described in Sect~2.1, KIC\,11285625 and
KIC\,6352430 were used as prototype stars for our simulations, and
the observed disentangled spectra of the KIC\,6352430,
KIC\,11285625, and KIC\,4931738 systems were used to test the code
on data of real binary systems. The results of the code application
to the simulated data are summarized in Table~\ref{Tab:
constrained_fitting}. Similar to the tests performed in Sect.~2.1,
we also applied the \gsspb\ module to the simulated spectra showing
anomalies in the abundances of certain chemical elements. The
results of this application are summarized in Table~\ref{Tab:
peculiarities_simulated}, in the columns indicated as ``\gsspb''.
The conclusions from our test runs on the simulated data are
essentially the same as in Sect.~2.1: i) the input parameters are
recovered in the majority of the cases, ii) the largest deviations
from the input values as well as the largest uncertainties are
observed for low S/N data, and iii) anomalies in the abundances of
chemical elements dominating the observed spectrum significantly
affect the determination of the global metallicity, which in turn
triggers deviations in other fundamental parameters of the star.
Since in the constrained fitting mode spectra of both binary
components are fitted simultaneously, any changes in the parameters
of one of the components will likely affect the parameter
determination for the companion star, due to changes in the light
dilution factor. The quality of the fit of the simulated data is
essentially the same as the one shown in Fig.~\ref{Fig:
KIC11285625_simulated}, for which reason it is not illustrated here.

\begin{figure}[t]
\includegraphics[scale=0.95]{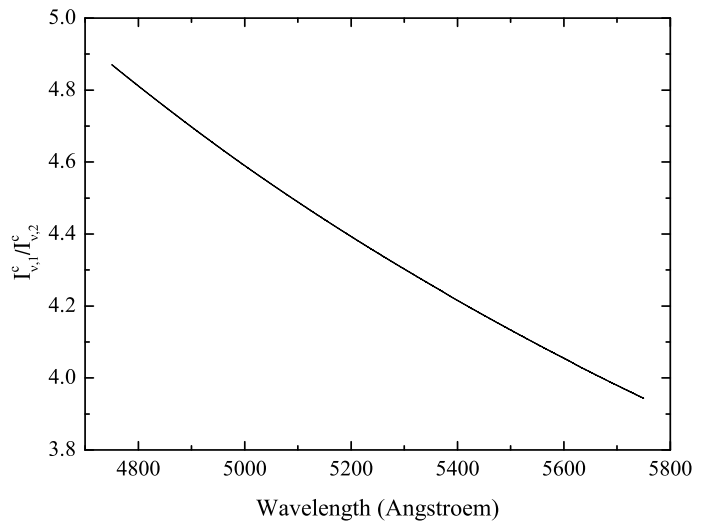}
\caption{{\small Ratio of the continuum intensity of the primary to
the one of the secondary in the KIC\,6352430 system. The models were
computed for the fundamental parameters listed in the last two
columns of Table~\ref{Tab: KIC6352430}}.} \label{Fig:
KIC6352430_contratio}
\end{figure}

The results of the application of the \gsspb\ software module to the
real spectra of KIC\,11285625, KIC\,6352430, and KIC\,4931738 are
summarized in Tables~\ref{Tab: KIC11285625}, \ref{Tab: KIC6352430},
and \ref{Tab: KIC4931738}, respectively (column indicated as
``\gsspb''). Significant deviations are observed for some of the
parameters from those obtained with the \gssps\ module
(unconstrained fitting) for the KIC\,11285625 system. This is likely
explained by the insufficient quality of the data for the secondary
component, whose parameter determination in turn impacts the
solution for the primary component. Indeed, the results of the
unconstrained fitting with the \gssps\ module (first two columns in
Table~\ref{Tab: KIC11285625}) suggest the sum of the dilution
factors to be well below unity. According to the definition of
disentangled spectra from Eq.~\ref{diluted_spectrum_gsspb}, this sum
is identical to unity in the constrained fitting mode, thus the
\gsspb\ code is in a sense forced to look for an alternative
solution that would satisfy the above mentioned criterion. The
change in the light dilution of each of the components in turn
triggers the differences in fundamental parameters compared to the
\gssps\ solution.

Certain deviations in the metallicity and effective temperature of
the secondary are observed for the KIC\,6352430 when compared to the
solution obtained with the \gssps\ module. Given that the latter
suggests unity for the sum of the light factors, the observed
discrepancies are likely due to wavelength dependence of the light
dilution factors. Indeed, the primary component is twice as hot as
its companion star, suggesting that the continuum intensity ratio
might vary significantly within the considered 1\,000~\AA\ wide
wavelength range. Figure~\ref{Fig: KIC6352430_contratio} illustrates
the variation of the ratio of the continuum intensity of the primary
to the one of the secondary as a function of wavelength, in the
range between 4\,750 and 5\,750~\AA. The ratio changes from
$\sim$4.9 at $\lambda\lambda$\ 4\,750~\AA\ to $\sim$3.9 at
$\lambda\lambda$\ 5\,750~\AA. For comparison, the variation in the
same quantity for the KIC\,11285625 system consisting of two similar
stars (see Table~\ref{Tab: KIC11285625}) does not exceed 0.2\% of
the value of 0.955 measured at 5\,250~\AA.

\subsection{\gsspc\ module}

The \gsspc\ software module was specifically designed for fitting
composite spectra of double-lined spectroscopic binary systems. We
refer the reader to Fig.~\ref{Fig: gsspb_diagram} and the
description there for details on the implemented algorithm as it is
very similar to the one implemented in the \gsspb\ module. The only
two differences in the \gsspc\ algorithm are: i) there is a
possibility to set the radial velocities of individual binary
components as free parameters, and ii) instead of comparing
theoretical spectra to the disentangled observed spectra of both
components, all possible combinations of synthetic primary and
secondary spectra from the computed grid are used to build composite
theoretical spectra of a binary which are then compared to the
observed spectrum on the scale of the latter. Composite theoretical
spectra are represented in the form
\begin{equation}
 R_{\rm comp}=\frac{I_{\nu,1}\Re_1^2 + I_{\nu,2}\Re_2^2}{I_{\nu,1}^c\Re_1^2 +
 I_{\nu,2}^c\Re_2^2}=\frac{\alpha R_1^{\rm th} + R_2^{\rm th}}{1+
 \alpha},
\end{equation}
where $\alpha$ is defined the same way as in
Eq.~\ref{diluted_spectrum_gsspb}.

We have tested the \gsspc\ module on simulated composite spectra of
binary stars, where the KIC\,11285625 and KIC\,6352430 binaries were
again used as the prototype systems. The results of our test runs
are summarized in Table~\ref{Tab: composite_fitting}; quality of the
fit to one of the spectra is illustrated in Fig.~\ref{Fig:
KIC11285625_simulated_composite}. As for the two previously
described software packages, we could recover the input parameters
for the majority of the models. However, we find that a combination
of low S/N and a small light contribution from either of the binary
components becomes a real bottleneck for fitting the composite
spectrum of a binary system. The only reasonable estimates in this
case are obtained for the radial velocities of stars, provided both
components show a sufficient amount of metal lines in their spectra
and the lines are well separated in velocity space. For the rest of
the parameters fitting the composite spectra fails, for which reason
there are no entries for the secondary components of KIC\,11285625
and KIC\,6352430 in Table~\ref{Tab: composite_fitting} at S/N of 60
and large values of the radii ratio. An assumption that one or both
stellar components of a binary show significant anomalies in their
spectra leads to the same results than have been obtained with the
\gsspb\ software package (see Sect.~2.2). For a limited amount of
models, we also performed test runs assuming that spectral lines of
individual spectral components overlap in the velocity space. To our
surprise, even in this case all initially assumed fundamental
parameters could be reasonably recovered, provided sufficiently high
S/N ($>$150) of the input data. No application of the \gsspc\ module
was done to real data of double-lined spectroscopic binaries,
however. The reason is that the individual composite observed
spectra of all three systems (KIC\,11285625, KIC\,6352430, and
KIC\,4931738) considered in our work have too low S/N for the method
to be robust.

\begin{figure}[t]
\includegraphics[scale=0.92]{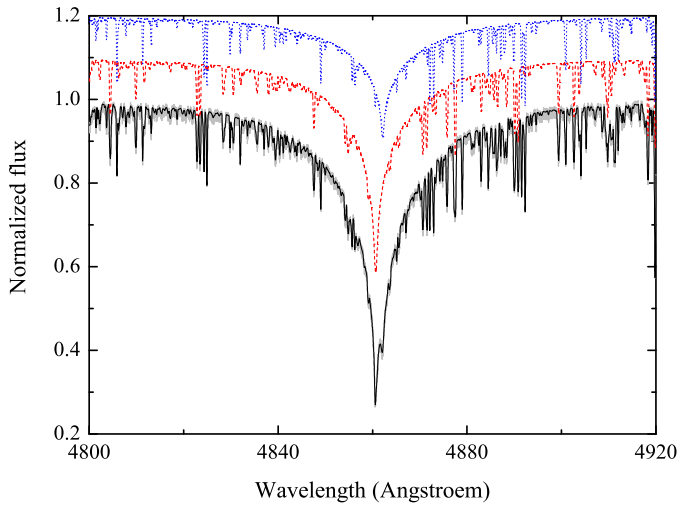}
\caption{{\small Quality of the fit of the simulated composite
spectrum of KIC\,11285625 (model 1a in Table~\ref{Tab:
composite_fitting}) with the \gsspc\ software package. The radii
ratio is 1.45, in agreement with the value reported by
\citet{Debosscher2013}. The simulated observed spectrum is shown
with a light grey line, the black solid line refers to the best fit
synthetic spectrum, the individual spectra of the primary/secondary
are shown by the red dashed/blue dotted lines. The individual,
RV-shifted spectra of the binary components were vertically shifted
for better visualization.}} \label{Fig:
KIC11285625_simulated_composite}
\end{figure}

The \gsspc\ software module is much more time consuming than the two
previously described modules. The reason is quite simple: assuming a
set up that includes calculation of $\sim$500 synthetic spectra, a
1\,000~\AA\ wide wavelength range, 8 CPUs, and three grid points in
RV for each of the components, the total number of the grid
combinations to be processed by the code is close to 1\,500\,000.
This is a factor of almost 10 more than for the same set up in the
\gsspb\ module, which transforms into about an hour of computation
time.

\section{Summary}

In this paper, we have presented a Fortran~90 code for spectrum
analysis of single stars and double-lined spectroscopic binary
systems. The code makes use of the {\sc OpenMPI} distribution and
runs in parallel mode.

There are three independent software modules included, each one is
designed for the application to a certain type of data. The \gssps\
module can be applied to the spectroscopic data of single stars as
well as to the disentangled, light diluted spectra of spectroscopic
binary components. In the latter case, the spectra of the stars are
interpreted independently of each other and the code assumes
wavelength-independent light dilution for each of the stars. The two
other modules, \gsspb\ and \gsspc, are specifically designed to fit
the spectra of double-lined spectroscopic binary systems. Where the
first module analyzes disentangled spectra, the second one focuses
on direct fitting of the composite spectra. The \gsspb\ module fits
disentangled spectra of both binary components simultaneously and
assumes that the light dilution factor of each of the components is
a function of wavelength. The light factors are replaced in the grid
by the ratio of the components' radii, which is obviously the same
for both stars. This way, an additional constraint is put on
individual light factors, namely their sum should equal unity at
each considered wavelength. The algorithm implemented in the \gsspc\
module is essentially the same, except that composite spectra are
fitted instead of disentangled ones. Radial velocities of the
individual stellar components can also be included into the fit if
necessary.

No matter which module is used for the spectroscopic analysis, the
user is strongly advised to start from reasonably coarse grid(s)
which would cover large parameter ranges. This step is absolutely
necessary to finally find a global minimum solution but not to be
stuck in a local minimum. For example, we advise to stick to the
range in the effective temperature that would cover the entire range
of spectral type in question. The full ranges for the surface
gravity and the metallicity currently available in the {\sc
LLmodels} library of atmosphere models (see Table~\ref{Tab: models}
for details) would be a good choice for the initial run. However,
the step sizes should be set to rather large values (e.g., 0.5 and
0.4 dex for \logg\ and [M/H, respectively]), to keep the calculation
time at a reasonable level. As soon as the first, coarse grid
solution is obtained, the user is advises to gradually narrow down
the parameter ranges, also making the grid finer. To speed up the
calculations with the \gsspc\ module, we would recommend to make a
priory estimate of the radial velocities of the individual binary
components (if at all possible) and to keep them fixed throughout
the analysis.

The \gssps\ and \gsspb\ software modules are quite fast and the
calculation time is reasonable even when the code is run on a quad
core desktop or laptop. If the disentangled spectra of both binary
components are available, and the difference in spectral types of
the two binary components is not large, we advise to start the
analysis with the unconstrained fitting mode implemented in the
\gssps\ module. This way, very good initial guesses can be obtained
for both stellar components of a binary system within short time.
With a good initial grid setup for the \gsspb\ module, the
physically more correct solution for both binary components can be
obtained within a limited amount of time. The disentangled spectra
of the components of a binary systems can be obtained in different
ways; the most widely used methodologies and codes are those
presented by \citet[][spectral disentangling in wavelength
domain]{Simon1994}, \citet{Hadrava1995} and \citet[][Fourier
spectral disentangling]{Ilijic2004}. The \gsspc\ module is more time
consuming than the two previous ones, thus we advise to use it when
i) very good quality data (S/N$>$150) are available, ii) good
initial guesses for the fitted parameters are provided, and iii) at
least some of the parameters can be fixed, e.g., from the light
curve solution.

Test runs of all three modules on the simulated data have shown that
all input parameters are well recovered in the majority of cases.
For the fitting of disentangled spectra, S/N$\geq$100 is likely to
be sufficient for decent analysis, whereas the composite spectrum
fitting generally requires a S/N above 150. The \gssp\ software
package is also suitable for the analysis of peculiar stars, though
one has to be aware of the fact that the global metallicity of the
stars can be significantly under/overestimated should the
peculiarity concern one of the dominant chemical elements in the
spectrum.

The application to real data has shown that for binary systems
consisting of two stars with significantly different spectral types,
the unconstrained fitting is likely to deliver not entirely correct
parameters. This is due to significant variations of the light
dilution factor with wavelength, as was the case for the
KIC\,6352430 system analyzed in this study.

In future, we plan to extend the software to the analysis of hot
stellar objects (early O-B spectral types) by including non-LTE
effects into the calculation of synthetic spectra. We will use the
so-called hybrid approach assuming non-LTE radiative transfer
calculations for chemical elements of interest, based on classical
LTE atmosphere models. The justification of such approach is
discussed in \citet{Nieva2007}. The implementation is rather
straightforward and consists of using pre-computed departure
coefficients from LTE in the {\sc SynthV} radiative transfer code.
The coefficients can be computed based on the LTE atmosphere models
with software such as the {\sc
tlusty\footnote{http://nova.astro.umd.edu/}} \citep{Hubeny1988}
code. We plan to perform such calculations for the entire hot part
of our grid of {\sc LLmodels} atmosphere models, so that non-LTE
spectrum synthesis can be done for any relevant model when it is
needed. Computing departure coefficients (including elements with
complex atomic structure) for a rather large sub-grid of atmosphere
models is a quite time-consuming procedure, which may take up to
several months of computation time. This prevents us from including
the non-LTE option in the current version of the software release.

\begin{acknowledgements}
The author is grateful to Prof. Vadim Tsymbal for providing the {\sc
SynthV} radiative transfer code and for the support in using it. The
whole IvS team and Dr. Holger Lehmann from Th\"{u}ringer
Landessternwarte Tautenburg are thanked for valuable discussions;
Dr. Peter Papics and Dr. Katrijn Clemer are acknowledged for careful
reading of the paper and providing a valuable feedback. The author
particularly acknowledges Prof. Dr. Conny Aerts for encouraging him
to develop the \gssp\ software package. Finally, the author wants to
thank the referee for careful reading of the manuscript and
providing valuable comments that helped to improve the paper. The
research leading to these results has received funding from the
European Community's Seventh Framework Programme FP7- SPACE-2011-1,
project number 312844 (SPACEINN).
\end{acknowledgements}

{}

\appendix

\section{Results of the analysis of simulated spectra}

\begin{table*}
\tabcolsep 1.5mm\caption{Results of the application of the \gssps\
software module to the simulated data of KIC\,11285625 (model 1) and
KIC\,6352430 (model 2) assuming different S/N ratio and dilution
factors. The ''Parameters of the input spectrum`` refer to the exact
parameter values assumed in the models. The error bars represent
1$\sigma$ uncertainties computed from $\chi^2$ statistics and taking
into account any possible correlations between different parameters.
The parameters are given in the form primary/secondary.}\label{Tab:
unconstrained_fitting}
\begin{tabular}{llllllll} \hline\hline
\multirow{2}{*}{model} & \multicolumn{1}{c}{\te\rule{0pt}{9pt}} & \multicolumn{1}{c}{\logg} & \multicolumn{1}{c}{[M/H]} & \multicolumn{1}{c}{$\xi$} & \multicolumn{1}{c}{\vsini} & \multicolumn{1}{c}{Dilution} & S/N\\
 & \multicolumn{1}{c}{(K)} & \multicolumn{1}{c}{(dex)} & \multicolumn{1}{c}{(dex)} & \multicolumn{1}{c}{(\kms)} & \multicolumn{1}{c}{(\kms)} & \multicolumn{1}{c}{factor} & \\\hline
\multicolumn{8}{c}{{\bf Parameters of the input spectrum\rule{0pt}{11pt}}}\\
\multicolumn{1}{c}{\multirow{2}{*}{1a}} & \multicolumn{1}{c}{\multirow{2}{*}{7\,000/7\,200}} & \multicolumn{1}{c}{\multirow{2}{*}{4.0/4.2}} & \multicolumn{1}{c}{\multirow{2}{*}{--0.2/+0.0}} & \multicolumn{1}{c}{\multirow{2}{*}{1.0/2.5}} & \multicolumn{1}{c}{\multirow{2}{*}{15/9}} & \multicolumn{1}{c}{\multirow{2}{*}{0.95/0.05}} & 60\\
 & & & & & & & 150\vspace{2mm}\\
\multicolumn{1}{c}{\multirow{2}{*}{1b}} & \multicolumn{1}{c}{\multirow{2}{*}{7\,000/7\,200}} & \multicolumn{1}{c}{\multirow{2}{*}{4.0/4.2}} & \multicolumn{1}{c}{\multirow{2}{*}{--0.2/+0.0}} & \multicolumn{1}{c}{\multirow{2}{*}{1.0/2.5}} & \multicolumn{1}{c}{\multirow{2}{*}{15/9}} & \multicolumn{1}{c}{\multirow{2}{*}{0.60/0.40}} & 60\\
 & & & & & & & 150\vspace{2mm}\\
\multicolumn{1}{c}{\multirow{2}{*}{1c}} & \multicolumn{1}{c}{\multirow{2}{*}{7\,000/7\,200}} & \multicolumn{1}{c}{\multirow{2}{*}{4.0/4.2}} & \multicolumn{1}{c}{\multirow{2}{*}{--0.2/+0.0}} & \multicolumn{1}{c}{\multirow{2}{*}{1.0/2.5}} & \multicolumn{1}{c}{\multirow{2}{*}{15/9}} & \multicolumn{1}{c}{\multirow{2}{*}{0.20/0.80}} & 60\\
 & & & & & & & 150\vspace{2mm}\\
\multicolumn{1}{c}{\multirow{2}{*}{2a}} & \multicolumn{1}{c}{\multirow{2}{*}{12\,750/6\,800}} & \multicolumn{1}{c}{\multirow{2}{*}{4.0/4.3}} & \multicolumn{1}{c}{\multirow{2}{*}{--0.1/--0.3}} & \multicolumn{1}{c}{\multirow{2}{*}{2.0/3.0}} & \multicolumn{1}{c}{\multirow{2}{*}{70/10}} & \multicolumn{1}{c}{\multirow{2}{*}{0.95/0.05}} & 60\\
 & & & & & & & 150\vspace{2mm}\\
\multicolumn{1}{c}{\multirow{2}{*}{2b}} & \multicolumn{1}{c}{\multirow{2}{*}{12\,750/6\,800}} & \multicolumn{1}{c}{\multirow{2}{*}{4.0/4.3}} & \multicolumn{1}{c}{\multirow{2}{*}{--0.1/--0.3}} & \multicolumn{1}{c}{\multirow{2}{*}{2.0/3.0}} & \multicolumn{1}{c}{\multirow{2}{*}{70/10}} & \multicolumn{1}{c}{\multirow{2}{*}{0.60/0.40}} & 60\\
 & & & & & & & 150\vspace{2mm}\\
\multicolumn{1}{c}{\multirow{2}{*}{2c}} & \multicolumn{1}{c}{\multirow{2}{*}{12\,750/6\,800}} & \multicolumn{1}{c}{\multirow{2}{*}{4.0/4.3}} & \multicolumn{1}{c}{\multirow{2}{*}{--0.1/--0.3}} & \multicolumn{1}{c}{\multirow{2}{*}{2.0/3.0}} & \multicolumn{1}{c}{\multirow{2}{*}{70/10}} & \multicolumn{1}{c}{\multirow{2}{*}{0.20/0.80}} & 60\\
 & & & & & & & 150\vspace{2mm}\\
\multicolumn{8}{c}{{\bf Results\rule{0pt}{11pt}}}\\
\multicolumn{1}{c}{\multirow{2}{*}{1a}} & 7\,020$^{+70}_{-66}$/7\,315$^{+490}_{-470}$\rule{0pt}{9pt} & 4.08$^{+0.08}_{-0.10}$/4.25$^{+0.55}_{-0.60}$ & --0.15$^{+0.07}_{-0.07}$/+0.13$^{+0.37}_{-0.34}$ & 1.12$^{+0.15}_{-0.15}$/2.77$^{+0.80}_{-0.78}$ & 15.2$^{+0.3}_{-0.3}$/8.7$^{+2.5}_{-2.5}$ & 0.93$^{+0.03}_{-0.03}$/0.05$^{+0.01}_{-0.01}$ & 60\vspace{2mm}\\
 & 7\,012$^{+30}_{-30}$/7\,200$^{+190}_{-200}$ & 3.99$^{+0.05}_{-0.05}$/4.14$^{+0.35}_{-0.35}$ & --0.19$^{+0.04}_{-0.04}$/+0.00$^{+0.15}_{-0.15}$ & 1.02$^{+0.10}_{-0.10}$/2.53$^{+0.55}_{-0.55}$ & 15.1$^{+0.2}_{-0.2}$/9.0$^{+1.2}_{-1.2}$ & 0.95$^{+0.02}_{-0.02}$/0.051$^{+0.005}_{-0.005}$ & 150\vspace{3mm}\\
\multicolumn{1}{c}{\multirow{2}{*}{1b}} & 7\,020$^{+60}_{-60}$/7\,215$^{+60}_{-60}$ & 4.00$^{+0.14}_{-0.14}$/4.13$^{+0.15}_{-0.15}$ & --0.16$^{+0.07}_{-0.07}$/+0.06$^{+0.09}_{-0.09}$ & 1.03$^{+0.17}_{-0.17}$/2.65$^{+0.18}_{-0.18}$ & 15.2$^{+0.4}_{-0.4}$/9.1$^{+0.4}_{-0.4}$ & 0.60$^{+0.02}_{-0.02}$/0.41$^{+0.02}_{-0.02}$ & 60\vspace{2mm}\\
 & 6\,995$^{+40}_{-40}$/7\,195$^{+45}_{-45}$ & 3.99$^{+0.06}_{-0.06}$/4.19$^{+0.07}_{-0.07}$ & --0.22$^{+0.03}_{-0.03}$/+0.00$^{+0.04}_{-0.04}$ & 1.02$^{+0.05}_{-0.05}$/2.52$^{+0.07}_{-0.07}$ & 15.1$^{+0.2}_{-0.2}$/9.1$^{+0.2}_{-0.2}$ & 0.61$^{+0.01}_{-0.01}$/0.41$^{+0.01}_{-0.01}$ & 150\vspace{3mm}\\
\multicolumn{1}{c}{\multirow{2}{*}{1c}} & 7\,070$^{+148}_{-158}$/7\,215$^{+50}_{-50}$ & 3.85$^{+0.40}_{-0.36}$/4.16$^{+0.09}_{-0.09}$ & --0.12$^{+0.15}_{-0.15}$/+0.01$^{+0.05}_{-0.05}$ & 1.19$^{+0.40}_{-0.40}$/2.57$^{+0.10}_{-0.10}$ & 15.2$^{+1.2}_{-1.2}$/8.9$^{+0.2}_{-0.2}$ & 0.20$^{+0.02}_{-0.02}$/0.79$^{+0.03}_{-0.03}$ & 60\vspace{2mm}\\
 & 7\,010$^{+60}_{-60}$/7\,200$^{+30}_{-30}$ & 3.98$^{+0.17}_{-0.17}$/4.20$^{+0.04}_{-0.04}$ & --0.16$^{+0.07}_{-0.07}$/+0.00$^{+0.03}_{-0.03}$ & 1.04$^{+0.16}_{-0.16}$/2.51$^{+0.04}_{-0.04}$ & 15.1$^{+0.5}_{-0.5}$/9.0$^{+0.2}_{-0.2}$ & 0.21$^{+0.01}_{-0.01}$/0.80$^{+0.01}_{-0.01}$ & 150\vspace{3mm}\\
\multicolumn{1}{c}{\multirow{2}{*}{2a}} & 12\,710$^{+163}_{-163}$/6\,870$^{+470}_{-455}$ & 4.03$^{+0.09}_{-0.09}$/4.37$^{+0.58}_{-0.62}$ & --0.08$^{+0.11}_{-0.11}$/--0.15$^{+0.36}_{-0.40}$ & 2.36$^{+0.69}_{-0.73}$/3.36$^{+1.20}_{-1.20}$ & 70.0$^{+6.0}_{-5.5}$/9.8$^{+1.2}_{-1.3}$ & 0.96$^{+0.03}_{-0.03}$/0.06$^{+0.02}_{-0.02}$ & 60\vspace{2mm}\\
 & 12\,750$^{+60}_{-60}$/6\,790$^{+176}_{-172}$ & 4.00$^{+0.04}_{-0.04}$/4.23$^{+0.46}_{-0.50}$ & --0.10$^{+0.05}_{-0.05}$/--0.30$^{+0.04}_{-0.04}$ & 2.03$^{+0.32}_{-0.32}$/3.03$^{+0.07}_{-0.07}$ & 70.1$^{+2.6}_{-2.6}$/10.1$^{+0.2}_{-0.2}$ & 0.95$^{+0.01}_{-0.01}$/0.05$^{+0.01}_{-0.01}$ & 150\vspace{3mm}\\
\multicolumn{1}{c}{\multirow{2}{*}{2b}} & 12\,860$^{+290}_{-300}$/6\,810$^{+60}_{-60}$ & 4.05$^{+0.13}_{-0.10}$/4.20$^{+0.17}_{-0.17}$ & --0.05$^{+0.16}_{-0.19}$/--0.25$^{+0.08}_{-0.08}$ & 2.55$^{+0.85}_{-0.95}$/3.15$^{+0.22}_{-0.22}$ & 69.5$^{+9.8}_{-8.5}$/10.1$^{+0.4}_{-0.4}$ & 0.60$^{+0.03}_{-0.03}$/0.41$^{+0.02}_{-0.02}$ & 60\vspace{2mm}\\
 & 12\,765$^{+125}_{-125}$/6\,800$^{+40}_{-40}$ & 4.00$^{+0.06}_{-0.06}$/4.29$^{+0.06}_{-0.06}$ & --0.09$^{+0.08}_{-0.08}$/--0.28$^{+0.07}_{-0.07}$ & 2.00$^{+0.40}_{-0.40}$/3.04$^{+0.12}_{-0.12}$ & 70.1$^{+4.3}_{-4.0}$/10.1$^{+0.7}_{-0.7}$ & 0.61$^{+0.01}_{-0.01}$/0.40$^{+0.01}_{-0.01}$ & 150\vspace{3mm}\\
\multicolumn{1}{c}{\multirow{2}{*}{2c}} & 12\,860$^{+800}_{-800}$/6\,810$^{+60}_{-60}$ & 4.05$^{+0.30}_{-0.30}$/4.26$^{+0.09}_{-0.09}$ & --0.01$^{+0.50}_{-0.50}$/--0.30$^{+0.06}_{-0.06}$ & 1.80$^{+2.50}_{-1.80}$/3.08$^{+0.11}_{-0.11}$ & 70.5$^{+15.0}_{-15.0}$/9.9$^{+0.4}_{-0.4}$ & 0.19$^{+0.02}_{-0.02}$/0.81$^{+0.02}_{-0.02}$ & 60\vspace{2mm}\\
 & 12\,800$^{+530}_{-530}$/6\,800$^{+40}_{-40}$ & 4.01$^{+0.15}_{-0.16}$/4.28$^{+0.05}_{-0.05}$ & --0.10$^{+0.22}_{-0.27}$/--0.30$^{+0.04}_{-0.04}$ & 1.91$^{+1.70}_{-1.70}$/3.02$^{+0.06}_{-0.06}$ & 70.2$^{+10.0}_{-10.0}$/10.0$^{+0.2}_{-0.2}$ & 0.20$^{+0.01}_{-0.01}$/0.80$^{+0.01}_{-0.01}$ & 150\vspace{1mm}\\
\hline
\end{tabular}
\end{table*}

\begin{table*}
\tabcolsep 1.8mm\caption{Same as Table~\ref{Tab:
unconstrained_fitting} but for the results of the application of the
\gsspb\ software module.}\label{Tab: constrained_fitting}
\begin{tabular}{llllllll} \hline\hline
\multirow{2}{*}{model} & \multicolumn{1}{c}{\te} & \multicolumn{1}{c}{\logg} & \multicolumn{1}{c}{[M/H]} & \multicolumn{1}{c}{$\xi$} & \multicolumn{1}{c}{\vsini} & \multicolumn{1}{c}{R$_1$/R$_2$} & S/N\\
 & \multicolumn{1}{c}{(K)} & \multicolumn{1}{c}{(dex)} & \multicolumn{1}{c}{(dex)} & \multicolumn{1}{c}{(\kms)} & \multicolumn{1}{c}{(\kms)} & & \\\hline
\multicolumn{8}{c}{{\bf Parameters of the input spectrum\rule{0pt}{11pt}}}\\
\multicolumn{1}{c}{\multirow{2}{*}{1a}} & \multicolumn{1}{c}{\multirow{2}{*}{7\,000/7\,200}} & \multicolumn{1}{c}{\multirow{2}{*}{4.0/4.2}} & \multicolumn{1}{c}{\multirow{2}{*}{--0.2/+0.0}} & \multicolumn{1}{c}{\multirow{2}{*}{1.0/2.5}} & \multicolumn{1}{c}{\multirow{2}{*}{15/9}} & \multicolumn{1}{c}{\multirow{2}{*}{1.45}} & 60\\
 & & & & & & & 150\vspace{2mm}\\
\multicolumn{1}{c}{\multirow{2}{*}{1b}} & \multicolumn{1}{c}{\multirow{2}{*}{7\,000/7\,200}} & \multicolumn{1}{c}{\multirow{2}{*}{4.0/4.2}} & \multicolumn{1}{c}{\multirow{2}{*}{--0.2/+0.0}} & \multicolumn{1}{c}{\multirow{2}{*}{1.0/2.5}} & \multicolumn{1}{c}{\multirow{2}{*}{15/9}} & \multicolumn{1}{c}{\multirow{2}{*}{2.50}} & 60\\
 & & & & & & & 150\vspace{2mm}\\
\multicolumn{1}{c}{\multirow{2}{*}{1c}} & \multicolumn{1}{c}{\multirow{2}{*}{7\,000/7\,200}} & \multicolumn{1}{c}{\multirow{2}{*}{4.0/4.2}} & \multicolumn{1}{c}{\multirow{2}{*}{--0.2/+0.0}} & \multicolumn{1}{c}{\multirow{2}{*}{1.0/2.5}} & \multicolumn{1}{c}{\multirow{2}{*}{15/9}} & \multicolumn{1}{c}{\multirow{2}{*}{4.00}} & 60\\
 & & & & & & & 150\vspace{2mm}\\
\multicolumn{1}{c}{\multirow{2}{*}{2a}} & \multicolumn{1}{c}{\multirow{2}{*}{12\,750/6\,800}} & \multicolumn{1}{c}{\multirow{2}{*}{4.0/4.3}} & \multicolumn{1}{c}{\multirow{2}{*}{--0.1/--0.3}} & \multicolumn{1}{c}{\multirow{2}{*}{2.0/3.0}} & \multicolumn{1}{c}{\multirow{2}{*}{70/10}} & \multicolumn{1}{c}{\multirow{2}{*}{0.50}} & 60\\
 & & & & & & & 150\vspace{2mm}\\
\multicolumn{1}{c}{\multirow{2}{*}{2b}} & \multicolumn{1}{c}{\multirow{2}{*}{12\,750/6\,800}} & \multicolumn{1}{c}{\multirow{2}{*}{4.0/4.3}} & \multicolumn{1}{c}{\multirow{2}{*}{--0.1/--0.3}} & \multicolumn{1}{c}{\multirow{2}{*}{2.0/3.0}} & \multicolumn{1}{c}{\multirow{2}{*}{70/10}} & \multicolumn{1}{c}{\multirow{2}{*}{0.90}} & 60\\
 & & & & & & & 150\vspace{2mm}\\
\multicolumn{1}{c}{\multirow{2}{*}{2c}} & \multicolumn{1}{c}{\multirow{2}{*}{12\,750/6\,800}} & \multicolumn{1}{c}{\multirow{2}{*}{4.0/4.3}} & \multicolumn{1}{c}{\multirow{2}{*}{--0.1/--0.3}} & \multicolumn{1}{c}{\multirow{2}{*}{2.0/3.0}} & \multicolumn{1}{c}{\multirow{2}{*}{70/10}} & \multicolumn{1}{c}{\multirow{2}{*}{1.40}} & 60\\
 & & & & & & & 150\vspace{2mm}\\
\multicolumn{8}{c}{{\bf Results\rule{0pt}{11pt}}}\\
\multicolumn{1}{c}{\multirow{2}{*}{1a}} & 7\,010$^{+90}_{-80}$/7\,280$^{+95}_{-130}$ & 3.95$^{+0.19}_{-0.20}$/4.30$^{+0.22}_{-0.26}$ & --0.16$^{+0.08}_{-0.08}$/+0.09$^{+0.09}_{-0.12}$ & 1.05$^{+0.21}_{-0.20}$/2.45$^{+0.22}_{-0.23}$ & 15.2$^{+0.5}_{-0.5}$/9.1$^{+0.6}_{-0.6}$ & 1.47$^{+0.11}_{-0.11}$ & 60\vspace{2mm}\\
 & 7\,000$^{+60}_{-60}$/7\,200$^{+60}_{-60}$ & 4.00$^{+0.06}_{-0.06}$/4.20$^{+0.08}_{-0.08}$ & --0.20$^{+0.04}_{-0.04}$/+0.01$^{+0.04}_{-0.04}$ & 0.98$^{+0.06}_{-0.06}$/2.50$^{+0.09}_{-0.08}$ & 15.1$^{+0.3}_{-0.3}$/9.0$^{+0.3}_{-0.3}$ & 1.45$^{+0.02}_{-0.02}$ & 150\vspace{3mm}\\
\multicolumn{1}{c}{\multirow{2}{*}{1b}} & 7\,020$^{+60}_{-60}$/7\,320$^{+208}_{-205}$ & 4.00$^{+0.14}_{-0.14}$/4.28$^{+0.45}_{-0.45}$ & --0.16$^{+0.07}_{-0.07}$/+0.16$^{+0.18}_{-0.20}$ & 1.03$^{+0.17}_{-0.17}$/2.59$^{+0.46}_{-0.43}$ & 15.2$^{+0.4}_{-0.4}$/8.8$^{+1.3}_{-1.3}$ & 2.72$^{+0.25}_{-0.25}$ & 60\vspace{2mm}\\\\
 & 7\,000$^{+50}_{-50}$/7\,200$^{+65}_{-65}$ & 4.00$^{+0.06}_{-0.06}$/4.20$^{+0.15}_{-0.16}$ & --0.20$^{+0.03}_{-0.03}$/+0.00$^{+0.06}_{-0.06}$ & 1.01$^{+0.06}_{-0.06}$/2.52$^{+0.18}_{-0.18}$ & 15.2$^{+0.3}_{-0.3}$/9.0$^{+0.5}_{-0.5}$ & 2.52$^{+0.14}_{-0.14}$ & 150\vspace{3mm}\\
\multicolumn{1}{c}{\multirow{2}{*}{1c}} & 7\,020$^{+60}_{-60}$/7\,320$^{+208}_{-205}$ & 4.00$^{+0.14}_{-0.14}$/4.28$^{+0.45}_{-0.45}$ & --0.16$^{+0.07}_{-0.07}$/+0.16$^{+0.18}_{-0.20}$ & 1.03$^{+0.17}_{-0.17}$/2.59$^{+0.46}_{-0.43}$ & 15.2$^{+0.4}_{-0.4}$/8.8$^{+1.3}_{-1.3}$ & 2.72$^{+0.25}_{-0.25}$ & 60\vspace{2mm}\\
 & 7\,015$^{+60}_{-60}$/7\,415$^{+460}_{-450}$ & 3.98$^{+0.11}_{-0.11}$/4.45$^{+1.20}_{-1.10}$ & --0.16$^{+0.06}_{-0.06}$/+0.15$^{+0.36}_{-0.37}$ & 1.05$^{+0.14}_{-0.16}$/3.30$^{+1.55}_{-1.46}$ & 15.2$^{+0.4}_{-0.4}$/8.1$^{+2.9}_{-3.4}$ & 4.94$^{+0.80}_{-1.14}$ & 150\vspace{3mm}\\
\multicolumn{1}{c}{\multirow{2}{*}{2a}} & 12\,590$^{+280}_{-280}$/6\,810$^{+70}_{-70}$ & 4.01$^{+0.09}_{-0.09}$/4.27$^{+0.17}_{-0.17}$ & --0.08$^{+0.22}_{-0.22}$/--0.28$^{+0.06}_{-0.06}$ & 2.23$^{+1.46}_{-1.43}$/3.05$^{+0.25}_{-0.25}$ & 72.2$^{+12.0}_{-12.0}$/10.2$^{+0.6}_{-0.6}$ & 0.51$^{+0.05}_{-0.05}$ & 60\vspace{2mm}\\
 & 12\,750$^{+125}_{-125}$/6\,800$^{+50}_{-50}$ & 4.00$^{+0.05}_{-0.05}$/4.29$^{+0.09}_{-0.09}$ & --0.12$^{+0.09}_{-0.09}$/--0.30$^{+0.04}_{-0.04}$ & 1.95$^{+0.60}_{-0.60}$/3.00$^{+0.10}_{-0.10}$ & 71.0$^{+4.9}_{-4.9}$/10.1$^{+0.3}_{-0.3}$ & 0.51$^{+0.02}_{-0.02}$ & 150\vspace{3mm}\\
\multicolumn{1}{c}{\multirow{2}{*}{2b}} & 12\,710$^{+225}_{-225}$/6\,880$^{+252}_{-255}$ & 4.04$^{+0.11}_{-0.11}$/4.58$^{+0.65}_{-0.70}$ & --0.08$^{+0.17}_{-0.17}$/--0.27$^{+0.09}_{-0.09}$ & 2.37$^{+0.95}_{-1.27}$/3.35$^{+0.49}_{-0.60}$ & 70.3$^{+8.9}_{-8.0}$/9.9$^{+1.3}_{-1.3}$ & 0.96$^{+0.16}_{-0.13}$ & 60\vspace{2mm}\\
 & 12\,770$^{+95}_{-90}$/6\,825$^{+76}_{-78}$ & 4.00$^{+0.05}_{-0.05}$/4.28$^{+0.17}_{-0.17}$ & --0.12$^{+0.08}_{-0.08}$/--0.30$^{+0.05}_{-0.05}$ & 1.95$^{+0.39}_{-0.39}$/3.02$^{+0.21}_{-0.21}$ & 72.3$^{+3.9}_{-3.9}$/9.9$^{+0.6}_{-0.6}$ & 0.91$^{+0.05}_{-0.05}$ & 150\vspace{3mm}\\
\multicolumn{1}{c}{\multirow{2}{*}{2c}} & 12\,665$^{+205}_{-195}$/6\,810$^{+445}_{-445}$ & 4.02$^{+0.10}_{-0.10}$/4.12$^{+0.85}_{-0.85}$ & --0.08$^{+0.15}_{-0.15}$/--0.30$^{+0.30}_{-0.30}$ & 2.29$^{+1.00}_{-1.02}$/3.33$^{+1.25}_{-1.00}$ & 70.0$^{+5.0}_{-5.0}$/9.9$^{+2.8}_{-2.8}$ & 1.40$^{+0.22}_{-0.22}$ & 60\vspace{2mm}\\
 & 12\,775$^{+88}_{-93}$/6\,790$^{+150}_{-150}$ & 3.98$^{+0.05}_{-0.05}$/4.24$^{+0.40}_{-0.33}$ & --0.11$^{+0.07}_{-0.07}$/--0.31$^{+0.15}_{-0.15}$ & 2.00$^{+0.39}_{-0.41}$/3.02$^{+0.45}_{-0.42}$ & 70.2$^{+3.5}_{-3.5}$/10.1$^{+1.2}_{-1.2}$ & 1.40$^{+0.08}_{-0.08}$ & 150\vspace{1mm}\\
\hline
\end{tabular}
\end{table*}

\begin{table*}
\tabcolsep 0.7mm\caption{Same as Table~\ref{Tab:
unconstrained_fitting} but for the results of the application of the
\gsspc\ software module.}\label{Tab: composite_fitting}
\begin{tabular}{lllllllll} \hline\hline
\multirow{2}{*}{model} & \multicolumn{1}{c}{\te} & \multicolumn{1}{c}{\logg} & \multicolumn{1}{c}{[M/H]} & \multicolumn{1}{c}{$\xi$} & \multicolumn{1}{c}{\vsini} & \multicolumn{1}{c}{RV} & \multicolumn{1}{c}{R$_1$/R$_2$} & S/N\\
 & \multicolumn{1}{c}{(K)} & \multicolumn{1}{c}{(dex)} & \multicolumn{1}{c}{(dex)} & \multicolumn{1}{c}{(\kms)} & \multicolumn{1}{c}{(\kms)} & \multicolumn{1}{c}{(\kms)} & &\\\hline
\multicolumn{9}{c}{{\bf Parameters of the input spectrum\rule{0pt}{11pt}}}\\
\multicolumn{1}{c}{\multirow{2}{*}{1a}} & \multicolumn{1}{c}{\multirow{2}{*}{7\,000/7\,200}} & \multicolumn{1}{c}{\multirow{2}{*}{4.0/4.2}} & \multicolumn{1}{c}{\multirow{2}{*}{--0.2/+0.0}} & \multicolumn{1}{c}{\multirow{2}{*}{1.0/2.5}} & \multicolumn{1}{c}{\multirow{2}{*}{15/9}} & \multicolumn{1}{c}{\multirow{2}{*}{--40/50}} & \multicolumn{1}{c}{\multirow{2}{*}{1.45}} & 60\\
 & & & & & & & & 150\vspace{2mm}\\
\multicolumn{1}{c}{\multirow{2}{*}{1b}} & \multicolumn{1}{c}{\multirow{2}{*}{7\,000/7\,200}} & \multicolumn{1}{c}{\multirow{2}{*}{4.0/4.2}} & \multicolumn{1}{c}{\multirow{2}{*}{--0.2/+0.0}} & \multicolumn{1}{c}{\multirow{2}{*}{1.0/2.5}} & \multicolumn{1}{c}{\multirow{2}{*}{15/9}} & \multicolumn{1}{c}{\multirow{2}{*}{--40/50}} & \multicolumn{1}{c}{\multirow{2}{*}{2.50}} & 60\\
 & & & & & & & & 150\vspace{2mm}\\
\multicolumn{1}{c}{\multirow{2}{*}{1c}} & \multicolumn{1}{c}{\multirow{2}{*}{7\,000/7\,200}} & \multicolumn{1}{c}{\multirow{2}{*}{4.0/4.2}} & \multicolumn{1}{c}{\multirow{2}{*}{--0.2/+0.0}} & \multicolumn{1}{c}{\multirow{2}{*}{1.0/2.5}} & \multicolumn{1}{c}{\multirow{2}{*}{15/9}} & \multicolumn{1}{c}{\multirow{2}{*}{--40/50}} & \multicolumn{1}{c}{\multirow{2}{*}{4.00}} & 60\\
 & & & & & & & & 150\vspace{2mm}\\
\multicolumn{1}{c}{\multirow{2}{*}{2a}} & \multicolumn{1}{c}{\multirow{2}{*}{12\,750/6\,800}} & \multicolumn{1}{c}{\multirow{2}{*}{4.0/4.3}} & \multicolumn{1}{c}{\multirow{2}{*}{--0.1/--0.3}} & \multicolumn{1}{c}{\multirow{2}{*}{2.0/3.0}} & \multicolumn{1}{c}{\multirow{2}{*}{70/10}} & \multicolumn{1}{c}{\multirow{2}{*}{--40/50}} & \multicolumn{1}{c}{\multirow{2}{*}{0.50}} & 60\\
 & & & & & & & & 150\vspace{2mm}\\
\multicolumn{1}{c}{\multirow{2}{*}{2b}} & \multicolumn{1}{c}{\multirow{2}{*}{12\,750/6\,800}} & \multicolumn{1}{c}{\multirow{2}{*}{4.0/4.3}} & \multicolumn{1}{c}{\multirow{2}{*}{--0.1/--0.3}} & \multicolumn{1}{c}{\multirow{2}{*}{2.0/3.0}} & \multicolumn{1}{c}{\multirow{2}{*}{70/10}} & \multicolumn{1}{c}{\multirow{2}{*}{--40/50}} & \multicolumn{1}{c}{\multirow{2}{*}{0.90}} & 60\\
 & & & & & & & & 150\vspace{2mm}\\
\multicolumn{1}{c}{\multirow{2}{*}{2c}} & \multicolumn{1}{c}{\multirow{2}{*}{12\,750/6\,800}} & \multicolumn{1}{c}{\multirow{2}{*}{4.0/4.3}} & \multicolumn{1}{c}{\multirow{2}{*}{--0.1/--0.3}} & \multicolumn{1}{c}{\multirow{2}{*}{2.0/3.0}} & \multicolumn{1}{c}{\multirow{2}{*}{70/10}} & \multicolumn{1}{c}{\multirow{2}{*}{--40/50}} & \multicolumn{1}{c}{\multirow{2}{*}{1.40}} & 60\\
 & & & & & & & & 150\vspace{2mm}\\
\multicolumn{9}{c}{{\bf Results\rule{0pt}{13pt}}}\\
\multicolumn{1}{c}{\multirow{2}{*}{1a}} & 7\,025$^{+100}_{-100}$/7\,175$^{+100}_{-100}$ & 4.02$^{+0.21}_{-0.21}$/4.11$^{+0.22}_{-0.22}$ & --0.12$^{+0.10}_{-0.10}$/+0.10$^{+0.12}_{-0.12}$ & 1.12$^{+0.25}_{-0.26}$/2.60$^{+0.27}_{-0.27}$ & 15.3$^{+0.6}_{-0.6}$/9.3$^{+0.7}_{-0.7}$ & --40.00$^{+0.28}_{-0.28}$/+50.00$^{+0.24}_{-0.22}$ & 1.38$^{+0.10}_{-0.10}$ & 60\vspace{2mm}\\
 & 6\,980$^{+40}_{-40}$/7\,210$^{+45}_{-45}$ & 3.99$^{+0.06}_{-0.06}$/4.17$^{+0.08}_{-0.08}$ & --0.19$^{+0.03}_{-0.03}$/+0.02$^{+0.04}_{-0.04}$ & 1.05$^{+0.09}_{-0.09}$/2.50$^{+0.09}_{-0.09}$ & 15.1$^{+0.3}_{-0.3}$/9.0$^{+0.2}_{-0.2}$ & --40.00$^{+0.12}_{-0.12}$/+50.00$^{+0.14}_{-0.14}$ & 1.45$^{+0.02}_{-0.02}$ & 150\vspace{3mm}\\
\multicolumn{1}{c}{\multirow{2}{*}{1b}} & \multicolumn{1}{c}{---} & \multicolumn{1}{c}{---} & \multicolumn{1}{c}{---} & \multicolumn{1}{c}{---} & \multicolumn{1}{c}{---} & \multicolumn{1}{c}{---} & \multicolumn{1}{c}{---} & 60\vspace{2mm}\\
 & 6\,990$^{+40}_{-40}$/7\,180$^{+60}_{-60}$ & 3.97$^{+0.06}_{-0.06}$/4.16$^{+0.16}_{-0.16}$ & --0.20$^{+0.02}_{-0.02}$/+0.01$^{+0.05}_{-0.05}$ & 1.00$^{+0.06}_{-0.06}$/2.66$^{+0.18}_{-0.18}$ & 15.1$^{+0.2}_{-0.2}$/8.9$^{+0.5}_{-0.5}$ & -39.95$^{+0.10}_{-0.10}$/+50.07$^{+0.16}_{-0.16}$ & 2.52$^{+0.06}_{-0.06}$ & 150\vspace{3mm}\\
\multicolumn{1}{c}{\multirow{2}{*}{1c}} & \multicolumn{1}{c}{---} & \multicolumn{1}{c}{---} & \multicolumn{1}{c}{---} & \multicolumn{1}{c}{---} & \multicolumn{1}{c}{---} & \multicolumn{1}{c}{---} & \multicolumn{1}{c}{---} & 60\vspace{2mm}\\
 & 6\,975$^{+40}_{-40}$/7\,190$^{+140}_{-140}$ & 3.97$^{+0.05}_{-0.05}$/4.12$^{+0.40}_{-0.40}$ & --0.21$^{+0.03}_{-0.03}$/+0.04$^{+0.13}_{-0.13}$ & 1.01$^{+0.05}_{-0.05}$/2.75$^{+0.50}_{-0.50}$ & 15.1$^{+0.2}_{-0.2}$/8.4$^{+1.3}_{-1.3}$ & -40.00$^{+0.11}_{-0.11}$/+50.00$^{+0.50}_{-0.50}$ & 4.11$^{+0.20}_{-0.20}$ & 150\vspace{3mm}\\
\multicolumn{1}{c}{\multirow{2}{*}{2a}} & 12\,990$^{+350}_{-350}$/6\,850$^{+160}_{-160}$ & 4.02$^{+0.15}_{-0.15}$/4.27$^{+0.22}_{-0.22}$ & --0.22$^{+0.25}_{-0.25}$/--0.30$^{+0.07}_{-0.07}$ & 2.77$^{+1.75}_{-1.71}$/3.04$^{+0.32}_{-0.32}$ & 73.6$^{+11.5}_{-11.5}$/9.8$^{+0.6}_{-0.6}$ & -40.0$^{+4.9}_{-4.9}$/+50.00$^{+0.15}_{-0.15}$ & 0.51$^{+0.04}_{-0.04}$ & 60\vspace{2mm}\\
 & 12\,760$^{+110}_{-110}$/6\,800$^{+35}_{-35}$ & 4.00$^{+0.04}_{-0.04}$/4.28$^{+0.07}_{-0.07}$ & --0.12$^{+0.08}_{-0.08}$/--0.27$^{+0.05}_{-0.05}$ & 2.09$^{+0.58}_{-0.58}$/3.01$^{+0.08}_{-0.08}$ & 70.6$^{+4.5}_{-4.5}$/9.8$^{+0.3}_{-0.3}$ & -40.05$^{+2.00}_{-1.90}$/+49.98$^{+0.12}_{-0.12}$ & 0.51$^{+0.03}_{-0.03}$ & 150\vspace{3mm}\\
\multicolumn{1}{c}{\multirow{2}{*}{2b}} & \multicolumn{1}{c}{---} & \multicolumn{1}{c}{---} & \multicolumn{1}{c}{---} & \multicolumn{1}{c}{---} & \multicolumn{1}{c}{---} & \multicolumn{1}{c}{---} & \multicolumn{1}{c}{---} & 60\vspace{2mm}\\
 & 12\,775$^{+110}_{-100}$/6\,810$^{+75}_{-75}$ & 3.97$^{+0.05}_{-0.05}$/4.28$^{+0.16}_{-0.16}$ & --0.16$^{+0.08}_{-0.08}$/--0.30$^{+0.05}_{-0.05}$ & 1.75$^{+0.50}_{-0.50}$/3.04$^{+0.20}_{-0.20}$ & 72.2$^{+4.2}_{-4.2}$/9.6$^{+0.7}_{-0.7}$ & -40.03$^{+2.19}_{-2.19}$/+50.00$^{+0.25}_{-0.25}$ & 0.91$^{+0.03}_{-0.03}$ & 150\vspace{3mm}\\
\multicolumn{1}{c}{\multirow{2}{*}{2c}} & \multicolumn{1}{c}{---} & \multicolumn{1}{c}{---} & \multicolumn{1}{c}{---} & \multicolumn{1}{c}{---} & \multicolumn{1}{c}{---} & \multicolumn{1}{c}{---} & \multicolumn{1}{c}{---} & 60\vspace{2mm}\\
 & 12\,775$^{+88}_{-93}$/6\,790$^{+150}_{-150}$ & 3.98$^{+0.05}_{-0.05}$/4.24$^{+0.40}_{-0.33}$ & --0.11$^{+0.07}_{-0.07}$/--0.31$^{+0.15}_{-0.15}$ & 2.00$^{+0.39}_{-0.41}$/3.02$^{+0.45}_{-0.42}$ & 70.2$^{+3.5}_{-3.5}$/10.1$^{+1.2}_{-1.2}$ & -40.00$^{+0.12}_{-0.12}$/+50.00$^{+0.14}_{-0.14}$ & 1.40$^{+0.08}_{-0.08}$ & 150\vspace{1mm}\\
\hline
\end{tabular}
\end{table*}

\begin{sidewaystable*}
\tabcolsep 2.5mm\caption{Results of the analysis of simulated
spectra of two binary systems, assuming peculiar abundances of Fe/Si
for the primary/secondary of KIC\,11285625, and of He for the
primary of KIC\,6352430. The model identification given in the
header is the same as in Tables~\ref{Tab: unconstrained_fitting},
\ref{Tab: constrained_fitting}, and \ref{Tab: composite_fitting},
where the input parameters can be read from. Asterisk refers to the
model with the metallicity fixed to the input value.}\label{Tab:
peculiarities_simulated}
\begin{tabular}{lllllllllllllll} \hline\hline
 \multirow{3}{*}{Parameter} & \multirow{3}{*}{Unit} &
 \multicolumn{4}{c}{\gsspb} & \multicolumn{4}{c}{\gssps} & \multicolumn{4}{c}{\gsspc}\\
  & & \multicolumn{2}{c}{KIC\,11285625} & \multicolumn{2}{c}{KIC\,6352430} & \multicolumn{2}{c}{KIC\,11285625} & \multicolumn{2}{c}{KIC\,6352430} & \multicolumn{2}{c}{KIC\,11285625} & \multicolumn{2}{c}{KIC\,6352430}\\
 & & \multicolumn{1}{c}{1b$^*$} & \multicolumn{1}{c}{1b} & \multicolumn{1}{c}{2b$^*$} & \multicolumn{1}{c}{2b}& \multicolumn{1}{c}{1b$^*$} & \multicolumn{1}{c}{1b} & \multicolumn{1}{c}{2b$^*$} & \multicolumn{1}{c}{2b} & \multicolumn{1}{c}{1b$^*$} & \multicolumn{1}{c}{1b} & \multicolumn{1}{c}{2b$^*$} & \multicolumn{1}{c}{2b}\\\hline
\multicolumn{14}{c}{{\bf Primary\rule{0pt}{11pt}}}\vspace{1mm}\\
\te & K & 7\,000$^{+50}_{-50}$ & 7\,030$^{+60}_{-60}$ & 12\,745$^{+115}_{-115}$ & 12\,780$^{+130}_{-130}$ & 7\,000$^{+30}_{-30}$ & 7\,025$^{+50}_{-50}$ & 12\,750$^{+110}_{-110}$ & 12\,750$^{+110}_{-110}$ & 6\,980$^{+30}_{-30}$ & 6\,985$^{+40}_{-40}$ & 12\,750$^{+100}_{-100}$ & 12\,750$^{+100}_{-100}$\vspace{1mm}\\
\logg & dex & 4.00$^{+0.05}_{-0.05}$ & 4.02$^{+0.11}_{-0.10}$ & 4.00$^{+0.05}_{-0.05}$ & 4.02$^{+0.06}_{-0.06}$ & 4.00$^{+0.05}_{-0.05}$ & 4.15$^{+0.15}_{-0.15}$ & 4.04$^{+0.05}_{-0.05}$ & 4.04$^{+0.05}_{-0.05}$ & 3.99$^{+0.04}_{-0.04}$ & 4.07$^{+0.09}_{-0.09}$ & 4.00$^{+0.04}_{-0.04}$ & 4.00$^{+0.04}_{-0.04}$\vspace{1mm}\\
$\xi$ & \kms & 0.96$^{+0.08}_{-0.08}$ & 0.82$^{+0.20}_{-0.20}$ & 2.01$^{+0.19}_{-0.21}$ & 2.45$^{+0.52}_{-0.55}$ & 0.96$^{+0.07}_{-0.07}$ & 0.63$^{+0.25}_{-0.25}$ & 2.00$^{+0.50}_{-0.50}$ & 2.00$^{+0.50}_{-0.50}$ & 1.01$^{+0.05}_{-0.05}$ & 0.97$^{+0.08}_{-0.08}$ & 2.07$^{+0.42}_{-0.42}$ & 2.07$^{+0.42}_{-0.42}$\vspace{1mm}\\
\vsini & \kms & 15.1$^{+0.2}_{-0.2}$ & 15.1$^{+0.2}_{-0.2}$ & 70.3$^{+4.0}_{-4.0}$ & 69.5$^{+4.0}_{-4.0}$ & 15.1$^{+0.2}_{-0.2}$ & 15.0$^{+0.3}_{-0.3}$ & 70.1$^{+4.3}_{-4.0}$ & 70.1$^{+4.3}_{-4.0}$ & 15.1$^{+0.2}_{-0.2}$ & 15.1$^{+0.2}_{-0.2}$ & 70.6$^{+3.6}_{-3.5}$ & 70.6$^{+3.6}_{-3.5}$\vspace{1mm}\\
${\rm [M/H]}$ & dex &--0.2$^*$ & --0.05$^{+0.07}_{-0.07}$ & --0.1$^*$ & --0.15$^{+0.09}_{-0.09}$ & --0.2$^*$ & 0.02$^{+0.12}_{-0.12}$ & --0.1$^*$ & --0.12$^{+0.12}_{-0.12}$ & --0.2$^*$ & --0.05$^{+0.06}_{-0.06}$ & --0.1$^*$ & --0.11$^{+0.07}_{-0.07}$ \vspace{1mm}\\
${\rm [El/H]}$ & dex & --4.48$^{+0.05}_{-0.05}$ & --4.45$^{+0.08}_{-0.08}$ & --1.30$^{+0.06}_{-0.07}$ & --1.31$^{+0.07}_{-0.07}$ & --4.48$^{+0.03}_{-0.03}$ & --4.42$^{+0.09}_{-0.09}$ & --1.29$^{+0.07}_{-0.10}$ & --1.29$^{+0.07}_{-0.10}$ & --4.48$^{+0.02}_{-0.02}$ & --4.45$^{+0.05}_{-0.05}$ & --1.29$^{+0.06}_{-0.07}$ & --1.29$^{+0.06}_{-0.07}$\vspace{1mm}\\
R$_1$/R$_2$ &  & 2.52$^{+0.05}_{-0.05}$ & 2.45$^{+0.12}_{-0.12}$ & 0.90$^{+0.04}_{-0.04}$ & 0.90$^{+0.05}_{-0.05}$ & 0.60$^{+0.01}_{-0.01}$ & 0.56$^{+0.07}_{-0.07}$ & 0.61$^{+0.01}_{-0.01}$ & 0.61$^{+0.01}_{-0.01}$ & 2.54$^{+0.06}_{-0.06}$ & 2.35$^{+0.08}_{-0.08}$ & 0.91$^{+0.04}_{-0.04}$ & 0.91$^{+0.04}_{-0.04}$\vspace{1mm}\\
RV & \kms & --- & --- & --- & --- & --- & --- & --- & --- & --40.02$^{+0.09}_{-0.09}$ & --39.98$^{+0.10}_{-0.10}$ & --40.06$^{+2.05}_{-1.95}$ & --40.03$^{+2.0}_{-2.0}$\\
\multicolumn{14}{c}{{\bf Secondary\rule{0pt}{11pt}}}\vspace{1mm}\\
\te & K & 7\,190$^{+70}_{-70}$ & 7\,185$^{+105}_{-105}$ & 6\,810$^{+80}_{-85}$ & 6\,800$^{+90}_{-90}$ & 7\,195$^{+40}_{-40}$ & 7\,195$^{+40}_{-40}$ & --- & --- & 7\,185$^{+65}_{-65}$ & 7\,190$^{+75}_{-75}$ & 6\,805$^{+80}_{-80}$ & 6\,805$^{+80}_{-80}$\vspace{1mm}\\
\logg & dex & 4.21$^{+0.14}_{-0.16}$ & 4.16$^{+0.36}_{-0.36}$ & 4.28$^{+0.17}_{-0.17}$ & 4.33$^{+0.16}_{-0.18}$ & 4.19$^{+0.06}_{-0.06}$ & 4.19$^{+0.06}_{-0.06}$ & --- & --- & 4.15$^{+0.16}_{-0.16}$ & 4.35$^{+0.17}_{-0.17}$ & 4.25$^{+0.16}_{-0.18}$ & 4.25$^{+0.16}_{-0.18}$\vspace{1mm}\\
$\xi$ & \kms & 2.52$^{+0.17}_{-0.17}$ & 2.58$^{+0.24}_{-0.39}$ & 3.02$^{+0.22}_{-0.20}$ & 2.96$^{+0.21}_{-0.25}$ & 2.52$^{+0.06}_{-0.06}$ & 2.52$^{+0.06}_{-0.06}$ & --- & --- & 2.64$^{+0.22}_{-0.22}$ & 2.05$^{+0.25}_{-0.24}$ & 3.07$^{+0.20}_{-0.20}$ & 3.07$^{+0.20}_{-0.20}$\vspace{1mm}\\
\vsini & \kms & 9.0$^{+0.5}_{-0.5}$ & 9.4$^{+0.7}_{-0.7}$ & 10.1$^{+0.5}_{-0.5}$ & 10.1$^{+0.5}_{-0.5}$ & 9.1$^{+0.2}_{-0.2}$ & 9.1$^{+0.2}_{-0.2}$ & --- & --- & 8.9$^{+0.5}_{-0.5}$ & 9.4$^{+0.7}_{-0.7}$ & 10.0$^{+0.5}_{-0.5}$ & 10.0$^{+0.5}_{-0.5}$\vspace{1mm}\\
${\rm [M/H]}$ & dex & 0.0$^*$ & --0.05$^{+0.09}_{-0.09}$ & --0.29$^{+0.08}_{-0.08}$ & --0.30$^{+0.10}_{-0.10}$ & 0.0$^*$ & 0.00$^{+0.03}_{-0.03}$ & --- & --- & 0.0$^*$ & --0.05$^{+0.10}_{-0.10}$ & --0.28$^{+0.08}_{-0.08}$ & --0.28$^{+0.08}_{-0.08}$\vspace{1mm}\\
${\rm [El/H]}$ & dex & --5.01$^{+0.35}_{-0.35}$ & --5.05$^{+0.52}_{-0.57}$ & --- & --- & --4.98$^{+0.14}_{-0.18}$ & --4.98$^{+0.14}_{-0.18}$ & --- & --- & --5.05$^{+0.34}_{-0.34}$ & --5.06$^{+0.40}_{-0.45}$ & --- & ---\vspace{1mm}\\
R$_1$/R$_2$ &  & 2.52$^{+0.05}_{-0.05}$ & 2.45$^{+0.12}_{-0.12}$ & 0.90$^{+0.04}_{-0.04}$ & 0.90$^{+0.05}_{-0.05}$ & 0.40$^{+0.01}_{-0.01}$ & 0.40$^{+0.01}_{-0.01}$ & --- & --- & 2.54$^{+0.06}_{-0.06}$ & 2.35$^{+0.08}_{-0.08}$ & 0.91$^{+0.04}_{-0.04}$ & 0.91$^{+0.04}_{-0.04}$\vspace{1mm}\\
RV & \kms & --- & --- & --- & --- & --- & --- & --- & --- & 49.98$^{+0.11}_{-0.11}$ & 50.00$^{+0.25}_{-0.25}$ & 50.00$^{+0.24}_{-0.22}$ & 50.05$^{+0.27}_{-0.25}$\vspace{1mm}\\
\hline
\end{tabular}
\end{sidewaystable*}

\clearpage
\onecolumn

\section{The \gssp\ package: Installation, running, and configurations files}

\subsection{Installation and running\vspace{2mm}}

The Grid Search in Stellar Parameters (\gssp) package is a
compilation of three separate modules, each one designed for the
analysis of a certain type of spectroscopic data. The code makes use
of an open source Message Passing Interface ({\sc OpenMPI})
implementation, which has to be installed along with the Intel
Fortran Compiler on the machine where the \gssp\ code is supposed to
run. The version of {\sc OpenMPI}~1.3.3 from July 14th, 2009 is
suitable; any stable version of the Intel Compiler is good but the
release should not be older than the one from the 12th of February,
2012 (version 12.1.3).

The code compiles with the following simple command:\vspace{2mm}

\texttt{mpif90 -o executable\_name *.f*}\vspace{2mm}\\
where``executable\_name'' is the name of an executable file (can be
anything the user likes). The compilation command is the same for
all the package modules. For the compilation to be successful, one
needs to run the compilation command twice. During the first run,
the GSSP\_module.f90 file is processed and the *.mod file with the
corresponding name is produced. This file contains the majority of
declarations as well as some other relevant information for other
subroutines. During the second run of the compilation command, the
compiler gets the information from the *.mod file and all remaining
Fortran source files get successfully processed.

To run the code, one need to use the following command:\vspace{2mm}

$\texttt{mpirun -n N [-hostfile Host\_file\_name] executable\_name
Input\_file\_name}\vspace{2mm}$\\
where \texttt{N} is the number of CPUs. The arguments given in
brackets are optional and not needed when the code is run on a
single machine with multiple-cores. The arguments become useful when
the code runs on a cluster PC with several nodes involved. The host
file contains information about the names of the individual nodes on
which the code will be running, and the number of CPUs on each of
those nodes. An example of such hostfile is given
below.\vspace{2mm}\\
node01\ \ \ \ \ slots=8\\
node02\ \ \ \ \ slots=8\\
node03\ \ \ \ \ slots=8\\
node04\ \ \ \ \ slots=8\\
node05\ \ \ \ \ slots=8\vspace{2mm}

\subsection{Configuration files\vspace{2mm}}

In this section, we discuss the structure of the input files for all
three modules of the \gssp\ package. Our intention is to give as
much detail as needed for each entry from those files.\vspace{3mm}

{\it \large The \gssps\ module}\vspace{3mm}\\
{\it Teff\_start\ \ \ Teff\_step\ \ \ Teff\_end}\ \ \ ! effective temperature range (K)\vspace{1mm}\\
{\it logg\_start\ \ \ logg\_step\ \ \ logg\_end}\ \ \ ! log surface gravity range (dex)\vspace{1mm}\\
{\it vmicro\_start\ \ \ vmicro\_step\ \ \ vmicro\_end}\ \ \ ! microturbulent velocity range (km\,s$^{-1}$)\vspace{1mm}\\
{\it vsini\_start\ \ \ vsini\_step\ \ \ vsini\_end}\ \ \ ! projected rotational velocity range (km\,s$^{-1}$)\vspace{1mm}\\
{\it dilution\_flag\ \ \ factor\_start\ \ \ factor\_step\ \ \ factor\_end}\ \ \ ! light dilution flag and light dilution factor range\vspace{1mm}\\
{\it abund\_flag\ \ \ [M/H]\_start\ \ \ [M/H]\_step\ \ \ [M/H]\_end}\ \ \ ! abundance flag and metallicity range (dex)\vspace{1mm}\\
{\it element\_id\ \ \ abund\_start\ \ \ abund\_step\ \ \ abund\_end}\ \ \ ! chemical element id (e.g., Fe or He) and abundance range\vspace{1mm}\\
{\it vmacro\ \ \ resolution}\ \ \ ! macroturbulent velocity (km\,s$^{-1}$) and resolving power\vspace{1mm}\\
{\it abundance\_path}\ \ \ ! absolute path to the abundance table(s)\vspace{1mm}\\
{\it model\_path}\ \ \ ! absolute path to the library of atmosphere models\vspace{1mm}\\
{\it vmicro\_model\ \ \ mass\_model}\ \ \ ! atmosphere model microturbulent velocity (km\,s$^{-1}$) and mass ($M_{\odot}$)\vspace{1mm}\\
{\it model\_comp\_flag}\ \ \ ! atmosphere model chemical composition flag (ST, CNm, or CNh)\vspace{1mm}\\
{\it nranges\ \ \ wave\_step\ \ \ mode}\ \ \ ! Number of wavelength regions, wavelength step (\AA), operational mode flag\vspace{1mm}\\
{\it spectrum\_path}\ \ \ ! Path to the observed spectrum\vspace{1mm}\\
{\it RV\_factor\ \ \ contin\_factor\ \ \ RV\ \ \ RV\_flag}\ \ \ ! RV scaling factor, continuum cutoff factor, RV value, RV option\vspace{1mm}\\
{\it wave\_start\ \ \ wave\_end}\ \ \ ! wavelength range(s) in \AA\vspace{1mm}\\
{\it wave\_start\ \ \ wave\_end}\vspace{3mm}\\
The first four entries in the configuration file set up the ranges
and step sizes for the effective temperature (\te), logarithm of
surface gravity (\logg), microturbulent velocity ($\xi$), and
projected
rotational velocity (\vsini). For example,\vspace{5mm}\\
6800\ \ 100\ \ 7200\ \ ! effective temperature range (K)\\
2.9\ \ 0.2\ \ 3.9\ \ ! log surface gravity range (dex)\\
2.8\ \ 0.4\ \ 4.4\ \ ! microturbulent velocity range (km\,s$^{-1}$)\\
35\ \ 5\ \ 50\ \ ! projected rotational velocity range
(km\,s$^{-1}$)\\

The next two entries set up the grid ranges for the light dilution
factor and the global metallicity [M/H] of the star. In both cases,
certain flags precede the parameter ranges. Should the light
dilution option be used in the calculations (in case of the analysis
of disentangled spectra of binary components), the value ``adjust''
must be set for the \texttt{dilution\_flag} parameter. Otherwise
(analysis of single stars), the value of the \texttt{dilution\_flag}
parameter needs to be set to ``skip''. Similarly, the
\texttt{abund\_flag} takes one of the two following values: ``skip''
or ``adjust''. The ``skip'' option assumes that the global
metallicity of the star can be optimized along with the other
fundamental parameters (\te, \logg, $\xi$, and \vsini). In this
case, individual abundances of all metals will be scaled by the same
amount. The ``adjust'' option assumes that the metallicity parameter
will be replaced in a grid by a chemical element (any, except
hydrogen) abundance. The individual abundances are optimized for one
chemical element at the time; the metallicity value [M/H] must
be fixed while all other fundamental parameters can be kept free. For example,\vspace{3mm}\\
adjust\ \ 0.6\ \ 0.1\ \ 1.0\ \ ! light dilution flag and light dilution factor range\\
skip\ \ -0.1\ \ 0.1\ \ 0.2\ \ ! abundance flag and metallicity range (dex)\vspace{3mm}\\
For detailed chemical composition analysis of the star, we recommend
to do the analysis in two steps: 1) optimizing the global
metallicity value together with the other four (and, optionally, the
light dilution factor) fundamental parameters; 2) fixing the
metallicity to the value derived in the first step and optimizing
individual abundances along with the other parameters.

The entry {\it element\_id\ \ abund\_start\ \ abund\_step\ \
abund\_end} is only relevant when the individual abundances have to be optimized. Below are the entry examples for Fe:\vspace{3mm}\\
Fe\ \ -4.59\ \ 0.05\ \ -4.39\ \ ! chemical element id (e.g., Fe or
He) and abundance range\vspace{3mm}\\
and for He\vspace{3mm}\\
He\ \ 0.0783\ \ 0.0005\ \ 0.0813\ \ ! chemical element id (e.g., Fe
or He) and abundance range\vspace{3mm}\\
The \texttt{element\_id} parameter refers to the element designation
(e.g., Fe for iron, He for helium); the numbers give the range and a
step width. Should the value of the \texttt{abund\_flag} parameter
be set to ``skip'', the entire entry will be ignored by the code.
{\bf Note:} though abundances of all metals are in logarithmic
scale, those of hydrogen and helium are not. Thus, when optimizing
the helium abundance, one has to remember that these are positive
values and the range should always be given from the smallest to the
largest value (see example above).

The {\it vmacro\ \ resolution} entry does not need any specific
remarks as it gives the values of macroturbulent velocity (in
km\,s$^{-1}$) and of the resolving power of the instrument. Both
parameters are used by the convolution code to add the corresponding
broadening to synthetic spectra. For example,\vspace{3mm}\\
0.0\ \ 32000\ \ ! macroturbulent velocity (km\,s$^{-1}$) and
resolving power\\

The next two entries, {\it abundance\_path} and {\it model\_path},
specify absolute paths to a file with individual abundances and to
the folder containing atmosphere models, respectively. For example:\vspace{3mm}\\
/home/UserName/Abundance\_table.abn\ \ ! absolute path to the abundance table(s)\\
/home/UserName/LLmodels/\ \ ! absolute path to the library of atmosphere models\vspace{3mm}\\
or\vspace{3mm}\\
/home/UserName/abundances/\ \ ! absolute path to the abundance table(s)\\
/home/UserName/LLmodels/\ \ ! absolute path to the library of atmosphere models\vspace{3mm}\\
A library of {\sc LLmodels} atmosphere models is included into the
the \gssp\ package; the library of interpolated Kurucz models with
the resolution matching the one of the {\sc LLmodels} grid is
provided upon request. The file with the individual abundances can
have any name but the extension *.abn is fixed. The program will not
run in the \texttt{abund\_flag = adjust} mode if the file name with
individual abundances is not provided. At the same time, the global
metallicity parameter [M/H] must be fixed when the name of *.abn
file is provided in the configuration file. Whenever [M/H] is set as
a free parameter, the path should be provided to the folder that
contains abundance tables corresponding to different global
metallicity values (e.g., /home/UserName/abundances/). These global
metallicity abundance tables have been pre-computed and are included
in the \gssp\ package distribution. Summarizing, the
\texttt{abundance\_path} parameter is closely connected to the [M/H]
parameter (whether it is free or not) and to the
\texttt{abund\_flag}:
\begin{itemize}
    \item {\bf [M/H] is a free parameter}: the absolute path to the folder
    with pre-computed chemical composition tables should be provided
    (e.g., /home/UserName/abundances/). Providing the name of a *.abn
    file in this case will result in an error message.
    \item {\bf [M/H] is fixed and \texttt{abund\_flag = skip}}: the
    other four fundamental parameters (\te, \logg, $\xi$, and
    \vsini) and (optionally) the dilution factor can be optimized based on the provided table with
    individual abundances. There are two options here: either one
    provides a path to the file that contains specific abundances
    (e.g., /home/UserName/Abundance\_table.abn), or a path to the folder
    containing pre-computed abundance tables corresponding to different metallicity values
    (e.g., /home/UserName/abundances/).
    \item {\bf [M/H] is fixed and \texttt{abund\_flag = adjust}}:
    the other four fundamental parameters (\te, \logg, $\xi$, and
    \vsini) and (optionally) the dilution factor can be optimized simultaneously with the individual
    abundances (one element at the time). In this case, a path to the file that contains specific elemental abundances has to be provided (e.g., /home/UserName/Abundance\_table.abn).
    Otherwise, the code will give an error message.
\end{itemize}

The next two entries in the configuration file, {\it vmicro\_model\
\ mass\_model} and {\it model\_comp\_flag}, set the values of the
atmosphere model microturbulent velocity (in km\,s$^{-1}$) and mass
($M_{\odot}$), and the value of the atmosphere model chemical composition flag. For example,:\vspace{3mm}\\
2\ \ 1\ \ ! atmosphere model microturbulent velocity (km\,s$^{-1}$) and mass ($M_{\odot}$)\\
ST\ \ ! atmosphere model chemical composition flag (ST, CNm, or CNh)\vspace{3mm}\\
All atmosphere models in the provided {\sc LLmodels} grid have been
computed for the fixed value of the microturbulent velocity of
2~km\,s$^{-1}$. However, the user is free to use his/her own grid of
models (in Kurucz format!) computed for a different value of the
microturbulence. The other two parameters, \texttt{mass\_model} and
\texttt{model\_comp\_flag}, were specifically included should the
user want to use {\sc marcs\footnote{http://marcs.astro.uu.se/}}
(Gustafsson et al. 2008) atmosphere models computed assuming a
different mass and/or chemical composition. Should the user choose
to use the provided grid of the {\sc LLmodels} atmosphere models,
both parameters must be fixed to the values given in the above
example. \textbf{Note:} when using spherical {\sc marcs} models, one
has to be aware that a hybrid approach is adopted in the spectrum
analysis -- spherical models and a plane-parallel radiative transfer
formalism in the spectrum synthesis code. The parameter
\texttt{model\_comp\_flag} takes one of the three following values:
``ST'' (stands for standard), ``CNm'' (stands for moderately
CN-cycled), or ``CNh'' (stands for heavily CN-cycled). The
``standard'' mixture reflects the typical elemental abundance ratios
in stars as a function of metallicity in the solar neighbourhood.
Two types of CN-cycled {\sc marcs} models reflect different carbon
isotopic ratios ($^{12}$C/$^{13}$C; see {\sc marcs} website for
details).

The {\it nranges\ \ wave\_step\ \ mode} entry sets the number of
wavelength ranges to be considered in the analysis, the wavelength
step width (in \AA), and the operational mode. For example:\vspace{3mm}\\
2\ \ 0.0376\ \ fit\ \ ! Number of wavelength regions, wavelength step (\AA), and operational mode flag\vspace{3mm}\\
The \texttt{mode} parameter can be assigned one of the following key
words: ``fit'' or ``grid''. The first, fitting mode, assumes that
the observed spectrum is provided (see below) and a grid of
synthetic spectra will be fitted to it. The quality of the fit is
evaluated based on the $\chi^2$-criterion; the $\chi^2$
distributions are provided for all free parameters as an output (see
below). The number of wavelength regions can be set to any integer
number, the ranges themselves are provided further in the
configuration file. The value of the step width in wavelength is
ignored in this case, and is computed directly from the observations
(see below). The second value that the \texttt{mode} parameter can
take is ``grid''. In this case, no fitting to the observations is
performed and a grid of synthetic spectra is computed in the
required parameter space. This option assumes that the spectra are
calculated in a single wavelength range, thus the number of
wavelength ranges (\texttt{nranges}) should be set to unity. The
step width in wavelength is also an important parameter in this
case, as it defines the wavelength grid for synthetic spectra.

The {\it spectrum\_path} entry provides a path (absolute or
relative) to the observed spectrum. This entry in the configuration
file will be ignored should the user run the code in the ``grid''
mode.
For example:\vspace{3mm}\\
home/UserName/Obs\_spectrum.dat\ \ ! Path to the observed spectrum\vspace{3mm}\\
The observed spectrum is an important part of the fitting mode, thus
an error message will be given when the spectrum is not provided but
the operational mode is set to ``fit''. The observed spectrum should
be provided in a two-column ASCII file, where the first and the
second columns refer to wavelength (in \AA, linear scale) and
normalized flux, respectively. \textbf{Note:} the wavelength scale
of the observed spectrum should be equidistant. The step width in
wavelength that will be used for the calculation of synthetic
spectra is computed from the observations.

The {\it RV\_factor\ \ \ contin\_factor\ \ \ RV\ \ \ RV\_flag} entry
is relevant when the fitting operational mode is chosen. Otherwise,
it will be ignored by the code. For example:\vspace{3mm}\\
0.50\ \ 0.99\ \ -35.5\ \ fixed ! RV scaling factor, continuum cutoff factor, RV value, RV option\vspace{3mm}\\
The \gssps\ module has an option to compute the cross-correlation
function (CCF) between the observations and the first synthetic
spectrum from the grid. The CCF will be computed when the
\texttt{RV\_flag} is assigned the key word ``adjust''. Otherwise
(\texttt{RV\_flag = fixed}), the code assumes that the radial
velocity of the observed spectrum is known and the RV value
preceding the \texttt{RV\_flag} in the configuration file
(-35.5~\kms\ in the example above) is used to correct the
observations. Should the CCF calculation be requested by the user,
the RV of the observed spectrum is determined as the first order
moment of the CCF. The velocity range in the CCF from which the
radial velocity is computed is defined by the user in a
parameterized way. The control parameter is \texttt{RV\_factor}
(takes the value of 0.5 in the example above), which sets the
intensity cut limit in the normalized to a unity CCF function. We
strongly recommend to use a value of 0.5 or slightly larger, which
in practice means that the CCF will be cut at its half maximum
intensity and the part above will be used to compute the first order
moment. Finally, the \texttt{contin\_factor} parameter takes a value
between 0 and 1 and reflects the cutoff value in the normalized
observed flux. All flux points above this value (0.99 in the example
above) will be taken into account in the calculation of the global
continuum correction factor. The information about the global
continuum position is taken from the synthetic spectra; the factor
is computed by means of the least-squares fit of the observations to
the theoretical spectrum. In practice, the value of this parameter
should be fairly large (between 0.95 and 1.0) to ensure that mainly
the continuum points are taken into account and the result is not
biased by the inclusion of weak spectral lines. Should the user not
want to do any continuum correction, the parameter can be set to
some unreliably large value (e.g., 10\,000), which will force the
correction factor to be identical to unity.

The {\it wave\_start\ \ wave\_end} entry refers to the wavelength
range to be considered in the analysis. The number of entries should
be equal to the number of wavelength regions (\texttt{nranges})
specified earlier in the configuration file (see above). For example:\vspace{3mm}\\
4750\ \ 5000\ \ ! wavelength range(s) in \AA\\
5200\ \ 5700\vspace{3mm}\\
Only one wavelength range can be provided when the operational mode
is set to ``grid''. In the fitting mode, the number of wavelength
regions is not limited, which makes the fitting of individual lines
in the spectrum possible.\vspace{3mm}

{\it \large The \gsspb\ module}\vspace{3mm}\\
{\it number\_components}\ \ \ ! number of stellar components\vspace{1mm}\\
{\it Teff\_start$^1$\ \ \ Teff\_step$^1$\ \ \ Teff\_end$^1$\ \ \ Teff\_start$^2$\ \ \ Teff\_step$^2$\ \ \ Teff\_end$^2$}\vspace{1mm}\\
{\it logg\_start$^1$\ \ \ logg\_step$^1$\ \ \ logg\_end$^1$\ \ \ logg\_start$^2$\ \ \ logg\_step$^2$\ \ \ logg\_end$^2$}\vspace{1mm}\\
{\it vmicro\_start$^1$\ \ \ vmicro\_step$^1$\ \ \ vmicro\_end$^1$\ \ \ vmicro\_start$^2$\ \ \ vmicro\_step$^2$\ \ \ vmicro\_end$^2$}\vspace{1mm}\\
{\it vsini\_start$^1$\ \ \ vsini\_step$^1$\ \ \ vsini\_end$^1$\ \ \ vsini\_start$^2$\ \ \ vsini\_step$^2$\ \ \ vsini\_end$^2$}\vspace{1mm}\\
{\it abund\_flag$^1$\ \ \ [M/H]\_start$^1$\ \ \ [M/H]\_step$^1$\ \ \ [M/H]\_end$^1$\ \ \ abund\_flag$^2$\ \ \ [M/H]\_start$^2$\ \ \ [M/H]\_step$^2$\ \ \ [M/H]\_end$^2$}\vspace{1mm}\\
{\it element\_id$^1$\ \ \ abund\_start$^1$\ \ \ abund\_step$^1$\ \ \ abund\_end$^1$\ \ \ element\_id$^2$\ \ \ abund\_start$^2$\ \ \ abund\_step$^2$\ \ \ abund\_end$^2$}\vspace{1mm}\\
{\it radii\_ratio\_start\ \ \ radii\_ratio\_step\ \ \ radii\_ratio\_end}\ \ \ ! ratio of the components' radii\vspace{1mm}\\
{\it vmacro$^1$\ \ \ vmacro$^2$\ \ \ resolution}\vspace{1mm}\\
{\it abundance\_path$^1$}\vspace{1mm}\\
{\it abundance\_path$^2$}\vspace{1mm}\\
{\it model\_path}\vspace{1mm}\\
{\it vmicro\_model\ \ \ mass\_model}\vspace{1mm}\\
{\it model\_comp\_flag}\vspace{1mm}\\
{\it nranges}\ \ \ ! Number of wavelength regions to be considered in the analysis\vspace{1mm}\\
{\it spectrum\_path$^1$}\ \ \ ! Path to the disentangled observed spectrum of the primary\vspace{1mm}\\
{\it spectrum\_path$^2$}\ \ \ ! Path to the disentangled observed spectrum of the secondary\vspace{1mm}\\
{\it RV\_factor$^1$\ \ \ RV$^1$\ \ \ RV\_flag$^1$\ \ \ contin\_factor$^1$\ \ \ RV\_factor$^2$\ \ \ RV$^2$\ \ \ RV\_flag$^2$\ \ \ contin\_factor$^2$}\vspace{1mm}\\
{\it wave\_start\ \ \ wave\_end}\vspace{1mm}\\
{\it wave\_start\ \ \ wave\_end}\vspace{3mm}\\
Since the structure of the configuration file itself and the meaning
of the input parameters are very similar to those of the \gssps\
module, we only discuss extra parameters specific to the \gsspb\
software module. In the above example of the configuration file,
superscripts ``1'' and ``2'' refer to the primary and secondary
components of a binary, respectively.

The {\it number\_components} refers to the number of stellar
components in the system. Currently, the value of this parameter
always has to be set to 2.

{\it radii\_ratio\_start\ \ \ radii\_ratio\_step\ \ \
radii\_ratio\_end} is another entry in the configuration file that
needs some attention. The entry refers to the binary components'
radii ratio and is obviously the same for both stellar components.
Depending on the binary system considered, this ratio can have
values close to one or above ten.\vspace{3mm}

{\it \large The \gsspc\ module}\vspace{3mm}\\
{\it number\_components}\ \ \ ! number of stellar components\vspace{1mm}\\
{\it Teff\_start$^1$\ \ \ Teff\_step$^1$\ \ \ Teff\_end$^1$\ \ \ Teff\_start$^2$\ \ \ Teff\_step$^2$\ \ \ Teff\_end$^2$}\vspace{1mm}\\
{\it logg\_start$^1$\ \ \ logg\_step$^1$\ \ \ logg\_end$^1$\ \ \ logg\_start$^2$\ \ \ logg\_step$^2$\ \ \ logg\_end$^2$}\vspace{1mm}\\
{\it vmicro\_start$^1$\ \ \ vmicro\_step$^1$\ \ \ vmicro\_end$^1$\ \ \ vmicro\_start$^2$\ \ \ vmicro\_step$^2$\ \ \ vmicro\_end$^2$}\vspace{1mm}\\
{\it vsini\_start$^1$\ \ \ vsini\_step$^1$\ \ \ vsini\_end$^1$\ \ \ vsini\_start$^2$\ \ \ vsini\_step$^2$\ \ \ vsini\_end$^2$}\vspace{1mm}\\
{\it abund\_flag$^1$\ \ \ [M/H]\_start$^1$\ \ \ [M/H]\_step$^1$\ \ \ [M/H]\_end$^1$\ \ \ abund\_flag$^2$\ \ \ [M/H]\_start$^2$\ \ \ [M/H]\_step$^2$\ \ \ [M/H]\_end$^2$}\vspace{1mm}\\
{\it element\_id$^1$\ \ \ abund\_start$^1$\ \ \ abund\_step$^1$\ \ \ abund\_end$^1$\ \ \ element\_id$^2$\ \ \ abund\_start$^2$\ \ \ abund\_step$^2$\ \ \ abund\_end$^2$}\vspace{1mm}\\
{\it radii\_ratio\_start\ \ \ radii\_ratio\_step\ \ \ radii\_ratio\_end}\ \ \ ! ratio of the components' radii\vspace{1mm}\\
{\it RV\_start$^1$\ \ \ RV\_step$^1$\ \ \ RV\_end$^1$\ \ \ RV\_start$^2$\ \ \ RV\_step$^2$\ \ \ RV\_end$^2$}\ \ \ ! radial velocity ranges (\kms)\vspace{1mm}\\
{\it vmacro$^1$\ \ \ vmacro$^2$\ \ \ resolution}\vspace{1mm}\\
{\it abundance\_path$^1$}\vspace{1mm}\\
{\it abundance\_path$^2$}\vspace{1mm}\\
{\it model\_path}\vspace{1mm}\\
{\it vmicro\_model\ \ \ mass\_model}\vspace{1mm}\\
{\it model\_comp\_flag}\vspace{1mm}\\
{\it nranges}\ \ \ ! Number of wavelength regions to be considered in the analysis\vspace{1mm}\\
{\it spectrum\_path}\ \ \ ! Path to the composite observed spectrum of the binary system\vspace{1mm}\\
{\it contin\_factor}\ \ \ ! continuum cutoff factor\vspace{1mm}\\
{\it wave\_start\ \ \ wave\_end}\vspace{1mm}\\
{\it wave\_start\ \ \ wave\_end}\vspace{3mm}\\
The meaning of the majority of the entries is the same as in the
\gssps\ and \gsspb\ modules, thus a detailed discussion is omitted
here.

{\it RV\_start$^1$\ \ \ RV\_step$^1$\ \ \ RV\_end$^1$\ \ \
RV\_start$^2$\ \ \ RV\_step$^2$\ \ \ RV\_end$^2$} is the only entry
that requires some attention as it is new compared to all previously
discussed. The entry obviously refers to the individual component
grids in radial velocities (in \kms). Should the user want to fix
the RV of either of the component stars, the \texttt{RV\_start}
parameter must be set equal to the \texttt{RV\_end} parameter.

\subsection{Output files\vspace{2mm}}

All output files are stored in the ``output\_files'' folder that is
located in the local work directory and is created automatically by
the code. The first important output file has the name
``ModelOverview.txt''. Whenever the user is out of the parameter
range provided by the library of atmosphere models, the code will
exit with an error message, referring the user to this file. It
gives an overview of all models that do (or do not) exist in the
specified parameter space. The file has the following structure:\\

[M/H]=-0.1\ \ Teff= 6800.0\ \ logg=2.90\ \ Vmicro=2.0\ \ Mass=1.0\ \
OK

[M/H]=-0.1\ \ Teff= 6800.0\ \ logg=3.10\ \ Vmicro=2.0\ \ Mass=1.0\ \
OK

[M/H]=-0.1\ \ Teff= 6800.0\ \ logg=3.30\ \ Vmicro=2.0\ \ Mass=1.0\ \
OK

[M/H]=-0.1\ \ Teff= 6800.0\ \ logg=3.50\ \ Vmicro=2.0\ \ Mass=1.0\ \
OK

[M/H]=-0.1\ \ Teff= 6800.0\ \ logg=3.70\ \ Vmicro=2.0\ \ Mass=1.0\ \
OK

$\ldots$

[M/H]=-1.0\ \ Teff= 6800.0\ \ logg=2.90\ \ Vmicro=2.0\ \ Mass=1.0\ \
Not found\vspace{3mm}\\
In this particular example, the code informs the user that one of
the low-metallicity atmosphere models is missing in the library.
Obviously, one has to reconsider the parameter space to stay within
the ranges of fundamental parameters assumed by the library, or to
consider using his/her own library of models that covers the
parameter range of interest.

Should the user request calculation of the CCF between the
observations and theoretical spectrum (\gssps\ and \gsspb\ modules),
the CCF itself will be saved in the ``output\_files/CCF.dat'' file.
This is a two-column ASCII file, with the columns 1 and 2 giving
radial velocity and the cross-correlation coefficient, respectively.
The radial velocity value determined as the first order moment of
the CCF is displayed on the screen. No file with the CCF is produced
by the \gsspc\ module as the RVs of both binary components are parts
of the grid search parameter space.

Each of the \gssp\ package modules produces output files that
contain the observations and the best fit synthetic spectra. These
are the files that have to be plotted to visually investigate the
quality of the fit. The output observed spectrum is corrected for
the RV shift (\gssps\ and \gsspb\ modules) and for possible
imperfections in global normalization. This is a two-column ASCII
file, where the first column gives the wavelength in \AA\ and the
second one gives the normalized flux. The best fit synthetic spectra
are also stored in ASCII files, and depending on the software module
used, can contain up to six columns. Below we give some guidance on
the content of the synthetic spectra output files:

\begin{itemize}
    \item {\bf \gssps\ module}: in the case of a single star spectrum fitting, the output file with the synthetic spectrum will have an extension *.rgs. The only
    relevant data for the user are in the first and the second columns, which give the wavelength in \AA\ and normalized flux, respectively.
    Should the user want to fit the disentangled spectrum of a binary
    component in the unconstrained mode (light dilution mode), the
    final light diluted synthetic spectrum is stored in a
    two-column ASCII file with *.rgs extension, which contains the
    wavelength in \AA\ (column 1) and the normalized diluted flux (column2).
    \item {\bf \gsspb\ module}: in the constrained fitting mode, the
    output best fit synthetic spectra files (one per stellar
    component) contain three columns: 1. the wavelength in \AA; 2.
    the normalized synthetic flux; 3. the normalized light diluted synthetic
    flux. Thus, columns 1 vs. 3 have to plotted for direct
    comparison with the observed disentangled spectra of the individual
    stellar components.
    \item {\bf \gsspc\ module}: in the case of the composite
    spectrum fitting, the best fit synthetic spectrum is stored in a
    four-column ASCII file. The meaning of the columns is the
    following: 1. the wavelength in \AA; 2. the composite normalized
    synthetic flux; 3. the normalized synthetic flux of the primary; 4.
    the normalized synthetic flux of the secondary. For direct
    comparison with the observations, the columns 1 vs. 2 have to be
    plotted.
\end{itemize}

Depending on the \gssp\ software module used, the output files
containing the spectra have slightly different names. In any case,
the file names speak for themselves and the observations can easily
be distinguished from the best fitting synthetic spectra.

The $\chi^2$ master file is saved in the
``output\_files/Chi2\_table.dat'' ASCII file. Depending on the
\gssp\ software module used, the file can contain up to 19 columns.
Below the headers of the $\chi^2$ output files from the different
\gssp\ modules are given.\vspace{2mm}

{\bf \gssps\ module, single star spectrum:\vspace{1mm}}

\hspace*{2mm}[M/H]\ \ \ \ \te\ \ \ \ \logg\ \ \ \ $\xi$\ \ \ \
[abundance] \ \ \ \ \vsini\ \ \ \ $\chi^2$\_inter\ \ \ \
contin\_factor\ \ \ \ reduced $\chi^2$\ \ \ \
$\chi^2_{1\sigma}$\vspace{2mm}

{\bf \gssps\ module, disentangled spectrum of a binary
components:\vspace{1mm}}

\hspace*{2mm}[M/H]\ \ \ \ \te\ \ \ \ \logg\ \ \ \ $\xi$\ \ \ \
[abundance] \ \ \ \ \vsini\ \ \ \ dilution\_factor \ \ \
$\chi^2$\_inter\ \ \ \ contin\_factor\ \ \ \ reduced $\chi^2$\ \ \ \
$\chi^2_{1\sigma}$\vspace{2mm}\\ The relevant columns for the user
are all fundamental parameters (and, optionally, dilution factor),
the reduced $\chi^2$, and the 1$\sigma$ level in $\chi^2$. The
parameter vs. reduced $\chi^2$ columns can be used to plot
individual $\chi^2$ distributions; the $\chi^2_{1\sigma}$ level can
be used to estimate 1$\sigma$ uncertainties. The parameter in
brackets is optional and will appear in the final table only if the
individual abundances are optimized. The $\chi^2$ master file is
sorted in a way that the best fit solution appears in the first row
of the file. The best fit parameters are also displayed on the
screen.\vspace{2mm}

{\bf \gsspb\ module:\vspace{1mm}}

\hspace*{2mm}[M/H]$_i$\ \ \ \ \te$_i$\ \ \ \ \logg$_i$\ \ \ \
$\xi_i$\ \ \ \ \vsini$_i$\ \ \ \ [abundance$_i$] \ \ \ \
(radii\_ratio) \ \ \ contin\_factor$_i$\ \ \ \ reduced $\chi^2_i$\ \
\ \ ($\chi^2_{\rm sum}$\ \ \ \ $\chi^2_{1\sigma}$)\vspace{2mm}

{\bf \gsspc\ module:\vspace{1mm}}

\hspace*{2mm}[M/H]$_i$\ \ \ \ \te$_i$\ \ \ \ \logg$_i$\ \ \ \
$\xi_i$\ \ \ \ \vsini$_i$\ \ \ \ [abundance$_i$] \ \ \ \ RV$_i$\ \ \
\ (radii\_ratio \ \ \ contin\_factor\ \ \ \ reduced $\chi^2$\ \ \ \
$\chi^2_{1\sigma}$)\vspace{2mm}\\ In the above two cases, the
subscript ``$i$'' refers to the stellar component (1 - for the
primary, 2 - for the secondary). The parameters given in parentheses
appear in the file only once and always follow the atmospheric
parameters of the secondary. In the above example of a \gsspb\
module output file, the parameters in the $\chi^2$ master file have
the following order: 1) the atmospheric parameters of the
primary+continuum correction factor and reduced $\chi^2$ for the
corresponding observed spectrum; 2) the atmospheric parameters of
the secondary, including the components' radii ratio+continuum
correction factor and reduced $\chi^2$ for the corresponding
observed spectrum; 3) the sum of the individual reduced $\chi^2$
values ($\chi^2_1+\chi^2_2$) and 1$\sigma$ level in $\chi^2_{\rm
sum}$.

{\bf Note:} when running the \gssps\ software module in the ``grid''
operational mode, the code produces an additional output folder
named ``rgs\_files''. This folder contains the entire grid of the
synthetic spectra.

\end{document}